\documentclass{article}

\usepackage{PRIMEarxiv}

\usepackage[utf8]{inputenc} 
\usepackage[T1]{fontenc}    
\usepackage{hyperref}       
\usepackage{url}            
\usepackage{booktabs}       
\usepackage{amsfonts}       
\usepackage{amsmath}
\usepackage{nicefrac}       
\usepackage{microtype}      
\usepackage{lipsum}
\usepackage{caption}
\usepackage{subcaption}
\usepackage{float}
\usepackage{fancyhdr}       
\usepackage{graphicx}       
\graphicspath{{media/}}     

\usepackage{eqnarray}
\usepackage{amsmath}
\usepackage{algorithm}
\usepackage{algpseudocode}
\usepackage{float}

\newtheorem{theorem}{Theorem}
\title{Genetic algorithm with a Bayesian approach for the detection of multiple points of change of time series of counting exceedances of specific thresholds
}

\author{
  Biviana Marcela Suárez-Sierra \\
  Computing and analytics area. School of Applied Sciences and Engineering. \\
  EAFIT University \\
  Medellín, Colombia\\
  \texttt{bmsuarezs@eafit.edu.co} \\
   \And
  Arrigo Coen \\
  Department of Mathematics, Faculty of Sciences. \\
  National Autonomous University of México \\
  CDMX, México, México\\
  \texttt{coen@ciencias.unam.mx} \\
  \And
  Carlos Alberto Taimal \\
  Computing and analytics area. School of Applied Sciences and Engineering. \\
  EAFIT University \\
  Medellín, Colombia\\
  \texttt{cataimaly@eafit.edu.co} \\
}

\begin{document}

\maketitle

\begin{abstract}

Although the applications of Non-Homogeneous Poisson Processes (NHPP) to model and study the threshold overshoots of interest in different time series of measurements have proven to provide good results, they needed to be complemented with an efficient and automatic diagnostic technique to establish the location of the change-points, which, when taken into account, make the estimated model fit poorly in regards of the information contained in the real one. Because of this, a new method is proposed to solve the segmentation uncertainty of the time series of measurements, where the generating distribution of exceedances of a specific threshold is the focus of investigation. One of the great contributions of the present algorithm is that all the days that trespassed are candidates to be a change-point, so all the possible configurations of overflow days under the heuristics of a genetic algorithm are the possible chromosomes, which will unite to produce new solutions. Also, such methods will be guarantee to non-local and the best possible one solution, reducing wasted machine time evaluating the least likely chromosomes to be a feasible solution. The analytical evaluation technique will be by means of the Minimum Description Length (\textit{MDL}) as the objective function, which is the joint posterior distribution function of the parameters of the NHPP of each regime and the change-points that determines them and which account as well for the influence of the presence of said times.

Thus, one of the practical implications of the present work comes in terms of overcoming the need of modeling the time series of measurements, where the distributions of exceedances of certain thresholds, or where the counting of certain events involving abrupt changes, is the main focus with applications in phenomena such as climate change, information security and epidemiology, to name a few.


\end{abstract}

\section{Introduction}\label{sec1}

The study of exceedances of a specific threshold in the time series of environmen\-tal measurements \cite{Rodrigues13},\cite{Suarez21}, data traffic measurements in computer security \cite{Hallgren2021} or the measurement of the prevalence of tuberculosis in New York \cite{Achcar2008} in certain time periods are some examples of the practicality of a model that allows us to establish whether certain exceedance mitigation contingencies in the measurements have worked correctly in an observed period, that is, if there are moments in the time series that indicate decrease, growth or steadiness in the rate of exceedance emission. These moments will be represented as abrupt changes, represented in the model as change-points. One of the most commonly used methodologies for this purpose is based on Non-Homogeneous Poisson Processes (NHPP), in which the change-points that determine the different regimes, which make up the entire time series, are estimated. In the same way, the parameters of the distribution of the excesses that compose each regime are estimated. For this, most of the time, the prior distribution of the change-points is considered to be the uniform distribution, with hyperparameters at the extremes of the interval where the researcher approximately observes the change-point. From there and from a Bayesian mechanism, the joint distribution of the change-points is updated with the rest of the parameters of each regime. The likelihood function of the observed overshootings and the aforementioned prior distribution are multiplied to obtain the posterior distribution. Using a Bayesian computational technique such as Markov Chain Monte Carlo (MCMC) on the posterior distribution, the parameters and change-points of interest are estimated.\\ 

It is worth noting that on the one hand we are talking about the time series of the measurements (TS1) and on the other hand about the time series of the number of times these measurements exceed a threshold of interest, so that cumulatively these counting measurements generate a new time series (TS2) of the number of exceedances at time $t$. The regimes and change points detected in TS2 will be the same as those determined in TS1. Since TS2 is a counting time series, it is possible to fit a Non-Homogeneous Poisson Process (NHPP) process, with a time-dependent overrun emission intensity function. This last feature gives the advantage to the model that it does not assume equal distribution between regimes, as conventional time series models do. Furthermore, regardless of the random mechanism that generates the TS1 measurements, by means of TS2, the number of change-points and the respective location in TS1 can be found. Because of the above, the objective function is organized over the TS2 observations in the likelihood and the parameters of the joint distribution function will be the same as those of the overshoot emission rate function which for the purposes of the present proposal, are functions with less than three parameters, as will be presented later in the expression (\ref{eq2}). That is, the complexity of the model will be directly related to the NHPP overshoot emission rate related to the TS2 observations with respect to time and not to the uncertainty of the TS1 series measurements.

One of the disadvantages that arise in the implementation of the aforementio\-ned models is the model execution time, because the more change-points are considered, the longer the time it will take to reach an optimum. But on the other hand, the greater the number of change-points, the greater the adjustment of the estimated to the observed, allowing less loss of information. The latter can be seen by comparing the Deviance information criterion (DIC) of the models with zero, one, two or more change-points in \cite{Suarez21}.  So we want a model that continues with the same goodness of fit, but automates the detection of change-points, in short run times. Therefore, in the present work we make a new proposal that unifies the principles of construction of the likelihood function from non-homogeneous Poisson processes, as it is a counting problem, and on the other hand, Bayesian computational methods to estimate the parameters of each regime and its change-points. After the joint posterior distribution is obtained, it will be combined with information theory principles to penalize the integer and real parameters present in the segmentation of the time series. All the above under the heuristics of a genetic algorithm, will convert the penalized posterior distribution into the objective function. Hence, the problem of solving the segmentation uncertainty with the present proposal is reduced to finding the penalized Maximum A Posteriori (MAP) estimator, which corresponds to a modification of the Minimum Description Length principle (MDL) and such we will refer to it as the Bayesian-MDL.

The cumulative mean function, which uniquely determines the NHPP, provides more information about the number of threshold exceedances up to a given time $t$. This mean function has an associated intensity function that will indicate the time trend in each of the segments and the speed of emission of the threshold overshoots of interest. The parameters of these functions will be estimated with the genetic algorithm, to which the starting conditions of the first generation of change-point configurations or chromosomes can be set. The latter will be responsible for generating the offspring with the best characteristics. To guarantee this, the mutation distribution, mating dynamics and reordering between chromosomes will be implemented and tested as in \cite{Li2012}. The execution time is also declared at the beginning of the algorithm and is determined by the number of generations. From each generation will come out the best of the children, which will be evaluated by the MDL, hence the chromosome with the lowest MDL among all the generations will be the best of all. In this type of algorithms, the execution time is compensated by the quality of the solution. Another advantage of this model is that once the cumulative mean function is known, the distribution of the number of overflows present at time $t$ can be explicitly established and from there different forecasting exercises can be performed.

It should also be established that as per the characterization exposed in \cite{Aminikhanghahi2017} for the type of method, the here proposed procedure is of the off-line type as the whole dataset is considered at once in a retrospective manner and the conditions set for the genetic algorithm are evaluated to determine an optimal solution. At the end of the algorithm execution, it will return time segments that will be independent and identically distributed (\textit{iid}) conditioned to the parameters of each segment. Assuming that the observations in each segment determined by the proposed algorithm are \textit{iid} may be wrong in the presence of correlation signals. However, \cite{Hallgren2021} comments that in the presence of outliers, the assumption turns out to be robust, as in the case of \cite{Fearnhead2006} and here.

The presentation of this work is structured as follows. In the second section, a background review will be presented related to what has been developed in the field of NHPP as methods of adjustment of the univariate function of cumulative mean of the exceedances of a threshold of interest up to time $t$ and its respective inferential analysis. In this same section the scheme of the genetic algorithm will be presented, displaying the start, pairing and stop conditions to guarantee that the solution is optimal. Section three will show the theoretical construction of the computational model for four different forms of the intensity function $\lambda(t)$ proposed in the literature. In section four we will present the results with three sets of simulated data and one set of real data. From the simulated data, the performance of the algorithm will be shown to determine the number of change-points, their location and the number of generations necessary to reach the optimal solution. With the real dataset, we will evaluate the ability of the algorithm to assess the impact of public policy on air pollution data in Bogotá by $PM_{2.5}$ between the years 2018 and 2020. Finally, in section five we will present the conclusions and future considerations.

\section{Inferential Analysis of the Univariate Non-Homogeneous Poisson Process}\label{sec2}

Based on the construction of the likelihood from a non-homogeneous Poisson process discussed in \cite{Cox66}, we have the following expression, from which we will start to construct the penalized MAP function. The developments that will be shown in the following sections will be based on the assumption that the true intensity function of the overshoots corresponds to some of the function in (\ref{eq2}). Then the process that we have correspond to a NHPP with intensities varying with respect to $t$,

\begin{eqnarray}
	\label{eq2}
	\begin{array}{llll}
		\lambda^{(W)}(t\mid\mathbf{\theta}) &=& (\alpha/\beta)(t/\beta)^{\alpha-1}, &\alpha ,\beta >0\\
		\lambda^{(MO)}(t\mid\mathbf{\theta}) &=& \frac{\beta}{t+\alpha}, &\alpha ,\beta >0\\
		\lambda^{(GO)}(t\mid\mathbf{\theta}) &=& \alpha\beta\exp (-\beta t), &\alpha ,\beta >0\\
		\lambda^{(GGO)}(t\mid\mathbf{\theta}) &=& \alpha\beta\gamma t^{\gamma-1}\exp (-\beta t^\gamma), &\alpha ,\beta , \gamma >0.\\
	\end{array}
\end{eqnarray}

Each letter in the superscript of the left-hand term corresponds to the initial letter of the name of the distribution used as intensity function as follows, Weibull ($W$) \cite{Govind95,Cid99}, Musa-Okumoto ($MO$) \cite{Musa84}, Goel-Okumoto ($GO$) and a generalization of the Goel-Okumoto model ($GGO$) \cite{Goel78}, for which the mean cumulative function, $m(t\mid\mathbf{\theta})$ is defined respectively by:

\begin{eqnarray}
	\label{eq1}
	\begin{array}{llll}
		m^{(W)}(t\mid\mathbf{\theta}) &=& (t/\beta)^{\alpha}, &\alpha ,\beta >0\\
		m^{(MO)}(t\mid\mathbf{\theta}) &=& \beta\log \left(1+\frac{t}{\alpha}\right), &\alpha ,\beta >0\\
		m^{(GO)}(t\mid\mathbf{\theta}) &=& \alpha [1-\exp (-\beta t)], &\alpha ,\beta >0\\
		m^{(GGO)}(t\mid\mathbf{\theta}) &=& \alpha [1-\exp (-\beta t^\gamma)], &\alpha ,\beta , \gamma >0.\\
	\end{array}
\end{eqnarray}

In the first three cases, the vector of parameters is $\theta=(\alpha ,\beta )$ and for the last one, $\theta=(\alpha ,\beta, \gamma)$.

The way the rate functions of the process are formulated shows that the observations within the same segment are equally distributed and independent, conditional on the parameters of the emission intensity function of the observa\-tions in that segment. Hence, this information is taken into account in the Bayesian-MDL to correspond to the growth, decrease or steadiness of excess emissions in each interval of the partition determined by the change-points. For example as exposed in \cite{Rodrigues13}, in air pollution models, the Weibull rate function shows a better fit than the GGO, without and with change-points.

The present model determines the number of change-points and their specific location over the time series of interest. Since the data suggested to be studied with this type of model are exceedances of a specific threshold, the nature of the change-points will represent an abrupt variation in the rate of growth or decay of the exceedances in the cumulative mean function, which in turn will represent variations in the original time series.

\subsection{Likelihood function with change-points}\label{sec2.1}

A change point is defined as an instant where the structural pattern of a time series shifts \cite{Aminikhanghahi2017,Truong20}. We assumed the presence of $J$ change-points, $\{\tau_1, \tau_2, \cdots ,\tau_J\}$ such that there are variations on the model parameters in between segments $\tau_{j-1} < t < \tau_j, j = 0, 1, 2, \ldots, J+1,  j_0 = 1, j_{J+1} = T$. These changes can be attributed to environmental policies or legislations in a certain year, the suspension of some network station due to maintenance for a climate case, macroeconomic policies from an economic standpoint or the presence of a stimulus in a neuroscience context. On the other side, assuming the overshoots from a given threshold between two regimes determined by the presence of a change-point can be modeled after a NHPP, then the intensity functions are,

\begin{eqnarray}
	\label{eq3}
\lambda(t \mid \theta) = 
\left\lbrace
\begin{array}{lll}
	\lambda(t\mid\theta_1), & 0\leq t< \tau_1, &\\
	\lambda(t\mid\theta_j), & \tau_{j-1}\leq t<\tau_j, & j = 2,3,\cdots,J,\\
	\lambda(t\mid\theta_{J+1}). &\tau_J\leq t\leq T,  &
\end{array}
\right.
\end{eqnarray}

where $\theta_j$ is the vector of parameters between the change-points $\tau_{j-1}$ and $\tau_j$, for $j=2,...,J$, and $\theta_1$ and $\theta_{J+1}$ are the parameter vectors before and after the first and last change-points, respectively.
With $n$ observations, the functions for the means are (see, e.g., \cite{Rodrigues13}),

\begin{eqnarray}
	\label{eq4}
m(t \mid \theta) = 
\left\lbrace
\begin{array}{lll}
	m(t\mid\theta_1), & 0\leq t< \tau_1, &\\
	m(\tau_1\mid\theta_1) + m(t\mid\theta_2) - m(\tau_1\mid\theta_2), & \tau_1\leq t<\tau_2, \\
	m(t\mid\theta_{j+1}) - m(\tau_j\mid\theta_{j+1}) \\+ \sum_{i=2}^{j}[m(\tau_i\mid\theta_i) - m(\tau_{i-1}\mid\theta_i)] \\+ m(\tau_1\mid\theta_1),  & \tau_{j}\leq t<\tau_{j+1}, & j = 2,3,\cdots,J,
\end{array}
\right.
\end{eqnarray}

where $\tau_{J+1}=T$. That is, because $m(t\mid\theta_1)$  represents the average number of exceedances of the standard, before the first change-point. $m(\tau_1\mid\theta_1) + m(t\mid\theta_2)-m(\tau_1\mid\theta_2)$ is the average number of exceedances of the standard between the first change-point $\tau_1$ and the second one $\tau_2$, given that the vector of parameters $\theta_2$ is known, and so on.

Be $D={d_1,\ldots ,d_n}$, where $d_k$ (as in the case without change-points), is the time of occurrence of the $k-th$ event (the $k-th$ time the maximum level of the environmental standard is exceeded, for example), with $k=1,2,...,n$, the likelihood function is determined by the expression below
where $N_{\tau_i}$ represents the exceedances before the change-point $\tau_i$, with $i=1,2,...,J$ (see \cite{Yang,Achcar11}).

\begin{align}
	\label{eq5}
	L(\mathbf{D}\mid\phi) \propto &  \left[\prod_{i=1}^{N_{\tau_1}}\lambda(d_i\ \mid \ \theta_1) \right]e^{-m(\tau_1 \mid \theta_1)}\nonumber\\
	& \times \left[ \prod_{j=2}^{J} \left( \prod_{i=N_{\tau_{j-1}}+1}^{N_{\tau_j}}\lambda(d_i \mid \theta_j)\right)e^{-[m(\tau_j \mid \theta_j)-m(\tau_{j-1} \mid \theta_j )]}  \right]\nonumber\\
	& \times \left[\prod_{i=N_{\tau_{J}}+1}^{n}\lambda(d_i\mid\theta_{J+1})\right]e^{-[m(T\mid\theta_{J+1})-m(\tau_{J}\mid\theta_{J+1} )]},
\end{align}

Using the expression (\ref{eq5}), we infer the parameters $\phi=(\theta, \tau)$, with $\theta=(\theta_1,...,\theta_J)$ and $\tau=(\tau_1,...,\tau_J)$ using a Bayesian approach. This perspective consists of finding the relationship between the prior distribution of the parameter $\theta$, on whose intensity function $\lambda(t\mid\theta)$ is dependent and the posterior distribution of the same, after taking into consideration the observed information $D$. In \cite{Achcar10}, this method was applied to obtained results very close to the observed ones, hence the descriptive capacity of the model and the methodology used. In such work, the criteria used to select the model that best fits the data together with the graphic part was the MDL.

\subsection{Detection of multiple change-points using genetic algorithm}\label{sec2.2}

\subsubsection{MDL framework}	

Since finding $J$ change-points implies finding out $J+1$ regimes for the time series or fitting $J+1$ models with different parameters, statiscal criteria has been used for such purpose in the available literature. Some include the Akaike Information Criterion (AIC), the Bayesian Information Criterion (BIC), Cross-Validation methods, and MDL-based methods. For problems involving regime shift detection, MDL methods usually provide superior empirical results. This superiority is probably due to the fact that both AIC and BIC apply the same penalty to all parameters, regardless of the nature of the parameter. On the other hand, MDL methods can adapt penalties to parameters depending on their nature be it continuous or discrete, bounded or not. In short, MDL defines the best fitting model as the one that enables the best compression of the data minimizing a penalized likelihood function. That is, 

\begin{align}
	\label{eq6}
MDL=-\log_2(L_{opt})+P.
\end{align}

Here $\log_2(L_{opt})$ is the required amount of information needed to store the fitted model, term taken from information theory. More details on this can be found in \cite{Davis2006}. $L_{opt}$ is obtained from replacing the maximum likelihood estimator in its function (\ref{eq5}). This will be explained in the next section.

Because of the above, it is possible to make the natural connection between the likelihood and the MDL objective function by means of the penalty $P$ (see \cite{Davis2006}). The broad penalty methodology is summarized in three principles as stated by \cite{Li2012}. The first one is to penalize the real valued parameters by the number of observations. Say $k$ that are used to estimate it, then, the penalty will be $\frac{\log_2 k}{2}$. For this principle, it is important to take into consideration how the observations are arranged to calculate the parameter of interest because this arrangement will be reflected in the penalty.

The second principle involves the penalty of how many integer parameters, such as the number of change-points $J$ and where they are located represented by $\tau_{1}, ..., \tau_{J}$ should be charged. This charging is calculated based on the value for each of them. For example, the quantity $J$, which is bounded by the total number of observations $T$ is charged an amount of $\frac{log_2T}{2}$. For each of the $\tau_{j}$ with $j=1,...,J$, we have that $\tau_{j}<\tau_{j+1}$, therefore the cost of its penalty will be $\frac{\log_2 \tau_{j+1}}{2}$ for $j=2,...,J$.

The last principle, mentioned in \cite{Li2012}, is the additivity principle. It involves constructing $P$ based on the sum of all the partial penalties mentioned above. The more parameters the model has, the higher $P$ will be. However, if despite adding parameters, the expression $\log_2(L_{opt})$ does not grow larger than the penalty $P$ of the extra parameters, the simpler model will be preferred. For the purposes of this paper, the following will be used as the
penalty function $P_{\tau}(\theta)$ for a fixed change point configuration,

\begin{equation}
\label{eq7}
P_{\tau}(\theta)=R\sum_{j=1}^{J+1} \frac{ln(\tau_1-\tau_{i-1})}{2} +ln(J) + \sum_{j=2}^{J}ln(\tau_j),    
\end{equation}

where $R=2,3$ depending on whether $\theta=(\alpha, \beta)$ or $\theta=(\alpha, \beta,\gamma)$, i.e. if $\theta$ has one or two parameters. The first summand of the right-hand term of expression represents that each of the real-valued parameters ($\alpha_j, \beta_j,\gamma_j$) will be penalized by $\frac{\ln(\tau_1-\tau_{j-1})}{2}$  of the $j$-th regime to which they belong and since there are $J+1$ regimes, the sum goes from 1 to $J+1$. The second summand of the right-hand term is the penalty of the number of points of change, and the last one comes from the sum of each of the penalties of each change-point.

\subsubsection{Genetic Algorithm Schema}
\label{Genetic Algorithm Scheme}

As exposed in \cite{Li2012} the total possible cases to evaluate the MDL corresponds to ${T\choose J}$, where $T$ is the number of observations in the time series and $J$ is the number of change-points. However, this number of parametric configurations is a quantity that does not make a computationally efficient optimization algorithm that aims to choose the best of the parametric configurations that minimize the MDL. For this reason, we will use the genetic algorithm that, by natural selection criteria will establish the best of the parameters configurations that we will  call chromosomes. Each chromosome will be labeled as $(J,\tau_{1},..., \tau_{J})$, where the first component $J$ stands for the number of change-points, located respectively at times $\tau_{1},...,\tau_{J}$, corresponding to the respective coordinates. The following is to establish how the genetic algorithm (GA) evaluates each of the chromosomes, while avoiding those probabilistically less optimal.

Let us now see how a complete generation is produced from an initial one with a given size, although the size can also be a couple. For this purpose, suppose there are $k$ individuals or chromosomes in the initial generation set at random. Each of the $T$ observations in the time series is allowed to be a change-point, independent of all other change-points, with probability, for example as seen in \cite{Li2012}, of 0.06. The number of change-points for each chromosome in the initial generation has a binomial distribution $Binomial (n = T-1, p = 0.06)$.

Two chromosomes are taken out of the initial generation, one mother and one father chromosome, to make a child of the next generation. This is done by a probabilistic combination of the parents. The principle of natural selection in this context will be performed by selecting a pair of chromosomes that best optimize the expression (\ref{eq6}) since this couple is considered to have the highest probability of having offspring. Therefore, the chromosomes are arranged from the most likely to the least one to have children, and each individual of the same generation is assigned a ranking, say $S_i$, being the ranking of the $j$ $th$ individual, with $S_j=1,...,k$. If $S_j=k$ then $j$ is the individual that best fits (\ref{eq6}) and if $S_j=1$ then $j$ is the individual that least well fits (\ref{eq6}).

Once this ranking has been made for each of the chromosomes of the same generation, we proceed to establish the probability of selection using the  following expression that simulates the principle of natural selection  of the parents that will generate the next generation.

\begin{equation}
\label{eq8}
\frac{S_j}{\sum_{i=1}^{k}S_i}    
\end{equation}

The chromosome that has the highest probability of being selected from the $k$ chromosomes, is chosen as the mother. Among the remaining $(k-1)$ chromosomes, the father is chosen under the same selection criteria as the mother. Suppose that the mother chromosome has $m$ change-points, located in some $\tau_1,\dots,\tau_m$, i.e. with the parameter configuration $(m,\tau_1,\dots,\tau_m)$. Similarly, suppose the father chromosome has the parametric configuration $(n,\delta_1,\dots,\delta_n)$. A child of these parents can arise simply by joining the two chromosomes, i.e., the child chromosome will initially have the following configuration, $(m+n,\epsilon_1,\dots,\epsilon_{m+n})$, where the $m+n$ change-points contain the mother's $m$ and the father's $n$ change-points.

After this, we remove the duplicated change-points from the child $(m+n,\epsilon_1, ..., \epsilon_{m+n})$. From this last configuration, we keep all or some change-points. For this, in  \cite{Li2012} use the dynamics of flipping a coin for each change-point in the child configuration. If heads comes up, the change-point is left, otherwise it is removed. That is, a binomial distribution will be used with probability parameter $1/2$ and number of trials, the length of the configuration of the child minus duplicities. All this with the aim that the offspring will keep traits of the parents, without being exact replicas.

Each point of change in the child chromosome, can undergo a mutation; a schema taken from \cite{Li2012} is that one of the following three alternatives may happen. We start by generating, with some random mechanism, the numbers $-1$, $0$ and, $1$ with respective probabilities $0.4$, $0.3$, $0.4$. If $-1$ comes out, the change point is subtracted by one unit; if $0$ comes out, it stays at the time it is at, and if  $1$ comes out, the current change point is added by one unit. Again, duplicates are eliminated. With this last procedure, we have finished the construction of Child 1. Child 2 up to $k$ are generated in the same way as the previous one.  New parents are selected if chromosomes are duplicated in the same generation with the previous parents.

The process of generation is repeated as many times as generations are to be obtained. In fact, one of the criteria for establishing the completion of the genetic algorithm is to fix the number of generations $r$.  Another approach could be to reach the solution that minimizes the objective function or one that does not improve after some given number of iterations or generations. Thus the objective function we
used was $\ln P_\tau(\theta) -  \ln f(D\mid\theta) - \ln f(\theta)$ and the optimization problem we intend to resolve through the means of the genetic algorithm takes the following form,

\begin{align}
	\label{eq9}
	\hat{\theta}_{BAYESIAN-MDL} = argmax_{\theta,\tau} \left(\ln P_\tau(\theta) -  \ln f_\tau(D\mid\theta) - \ln f_\tau(\theta)\right)
\end{align}

In other words, $\hat{\theta}_{BAYESIAN-MDL}$ is the maximum argument of the objective function, in which case it will be the optimal solution to the problem of finding the best configuration of shift points and the respective parameters of the regimes they determine.

\section{Methods and Models}\label{sec3}

We are particularly interested in the likelihood function of expression (\ref{eq9}) in order to establish what we have called the corresponding Bayesian-MDL. Therefore, for any $ m (t \mid \theta)$ defined in expression (\ref{eq1}) and its respective intensity function defined in (\ref{eq2}), it follows that,

\begin{align}\label{eq10}
\begin{split}
L(\mathbf{D}\mid\phi) &\propto e^{-m(\tau_1\mid\theta_1)} \prod_{i=1}^{N_{\tau_1}} \lambda(d_i\mid\theta_1)\\
&\times \prod_{j=2}^{J} \left( e^{-[m(\tau_j\mid\theta_j)-m(\tau_{j-1} \mid \theta_j )]}\prod_{i=N_{\tau_{j-1}}+1}^{N_{\tau_j}} \lambda(d_i \mid\theta_j) \right)\\
& \times e^{-[m(T \mid\theta_{J+1})-m(\tau_{J}\mid\theta_{J+1} )]} \prod_{i=N_{\tau_{J}}+1}^{n}\lambda(d_i\mid\theta_{J+1}) \\
\end{split}
\end{align}

since the product with respect to $i$ only affects the functions $\lambda(t,\theta)$ in each of the different $J+1$ regimes determined by the $J$ change-points in the vector $\tau =(\tau_1, \tau_2,\dots, \tau_j)$.
Using that for $J$ change-points, $\tau_{J+1} := T$ where $T$ is the number of daily measurements, $\tau_0=0,N_0=0$ and $m(0\mid\theta)=0$, it follows that the expression (\ref{eq10}) reduces to,

\begin{equation}
\label{eq11}
L(\mathbf{D}\mid\phi) \propto  \prod_{j=1}^{J+1} \left(e^{-[m(\tau_j\mid\theta_j)-m(\tau_{j-1} \mid \theta_j )]} \prod_{i=N_{\tau_{J-1}}+1}^{N_{\tau_j}} \lambda(d_i\mid\theta_j)   \right)  
\end{equation}

Taking the logarithm of (\ref{eq11}) we have,

\begin{align}%
	\label{eq12}
	\log L(D \mid \phi) &= \left( \sum_{j=1}^{J+1}   m(\tau_{j-1}\mid\theta_j)-m(\tau_j\mid\theta_j)   \right) \nonumber\\
	 &+\left(\sum_{j=1}^{J+1} \quad \sum_{i= N_{\tau_{j-1}}+1}^{ N_{\tau_j}} \log\lambda(d_i\mid\theta_j) \right)\nonumber\\
	&=\sum_{j=1}^{J+1} \left(  m(\tau_{j-1}\mid\theta_j)-m(\tau_j\mid\theta_j)+\sum_{i= N_{\tau_{j-1}}+1}^{ N_{\tau_j}} \log\lambda(d_i\mid\theta_j)   \right)
\end{align}

Also, the Bayesian principle states that,

\begin{align}%
	\label{eq13}
f(\theta\mid D)\propto L(D\mid\theta) f(\theta), 
\end{align}

where $L$ is the likelihood function depending on the observations $D$, given the parameter vector $\theta$ and, $f(\theta)$ the prior function of the parameter vector $\theta$. Taking logarithm for (\ref{eq13}) we have that,

\begin{align}%
	\label{eq14}
\ln f(\theta\mid D)\propto \ln(L(D\mid\theta)) + \ln(f(\theta))
\end{align}

From (\ref{eq14}), we start by establishing the form of the prior joint function $f(\theta)=f(\alpha, \beta, \tau_j)$ of the first three cases in (\ref{eq1}). The expression for the last case is given in the appendix (\ref{sec:DerivationObjFunctions}).

\subsection{Prior distributions}

If we take $\alpha \sim Gamma(\phi_{11},\phi_{12})$, then,
\[f(\alpha) = \frac{\phi_{11}^{\phi_{12}}}{\Gamma(\phi_{12})} \alpha^{\phi_{12}-1}e^{-\phi_{11} \alpha} \]

After applying logarithm we obtain,

\begin{align} \label{eq15}
	\log  f(\alpha) &= \log\left( \dfrac{\phi_{11}^{\phi_{12}}}{\Gamma(\phi_{12})} \alpha^{\phi_{12}-1}e^{-\phi_{11} \alpha} \right)\nonumber\\
	&=  \phi_{12}\log\phi_{11} - \log\Gamma(\phi_{12}) + (\phi_{12}-1)\log\alpha - \phi_{11}\alpha\nonumber\\
	&\propto (\phi_{12}-1)\log\alpha - \phi_{11}\alpha
\end{align}

Similarly for $\beta$, if we take $\beta \sim Gamma(\phi_{21},\phi_{22})$, then,
\begin{align} \label{eq16}
	\log f(\beta) \propto (\phi_{22}-1)\log\beta - \phi_{21}\beta
\end{align}

On the other hand, assuming every time in the series can be chosen as a change-point $\tau_j$ with the same probability, thus $\tau_j \sim Uniform(1, T), \ j = 1, 2, \ldots, J$.

Then we have,

\begin{align} \label{tau_unif}
	f(\tau_j) = \frac{1}{T-1}
\end{align}

Taking logarithm we obtain,

\begin{align} \label{tau_unif_log}
	\log f(\tau_j) = - log(T-1)
\end{align}

Rebuilding the joint function for $\theta=(\alpha,\beta, \tau_j)$ under the assumption of independence, we have,

\begin{align}
	\label{eq17}
	\begin{split}
\log f(\alpha,\beta, \tau_j) &\propto(\phi_{12}-1)\log\alpha - \phi_{11}\alpha \\
&+ (\phi_{22}-1)\log\beta - \phi_{21}\beta - log(T-1)
\end{split}
\end{align}

Up to this point, the second summand of the right-hand side of (\ref{eq14}) has been obtained. Next, the first summand of the right-hand side of (\ref{eq14}) will be derived. This will be done for the first case of (\ref{eq1}) and the other ones are given in the appendix (sections (\ref{sec:MODerivation}), (\ref{sec:GODerivation}), (\ref{sec:GGODerivation})).

\subsubsection{Weibull intensity rate (W)}
\label{Weibull}

After taking the expressions for the intensity function $\lambda^{(W)}(t\mid\theta)$ and the cumulative mean function $m^{(W)}(t\mid\theta)$ using (\ref{eq1}) and (\ref{eq2}) respectively, and replacing these in (\ref{eq12}) we have,
    
\begin{equation}
\begin{aligned}
	\label{eq19}
	\log L(D\mid\phi)&=\sum_{j=1}^{J+1} \left(  m(\tau_{j-1}\mid\theta_j)-m(\tau_j\mid\theta_j)+\sum_{i= N_{\tau_{j-1}}+1}^{ N_{\tau_j}} \log\lambda(d_i\mid\theta_j)   \right)\nonumber\\
	&=\sum_{j=1}^{J+1} \left(\left(\dfrac{\tau_{j-1}}{\beta_j}\right)^{\alpha_j}-\left(\dfrac{\tau_{j}}{\beta_j}\right)^{\alpha_j} \right. \\
	&\left. +\sum_{i= N_{\tau_{j-1}}+1}^{ N_{\tau_j}} \log\left(\dfrac{\alpha_j}{\beta_j} \left(\dfrac{d_i}{\beta_j}\right)^{\alpha_j-1}\right)\right)\nonumber\\
	&=\sum_{j=1}^{J+1} \left( \dfrac{\tau_{j-1}^{\alpha_j}-\tau_{j}^{\alpha_j}}{\beta_j^{\alpha_j}} + (N_{\tau_j}-N_{\tau_{j-1}})\left(\log(\alpha_j)- \alpha_j\log(\beta_j)\right)\right.\nonumber\\
	 &\left. +(\alpha_j-1)\sum_{i= N_{\tau_{j-1}}+1}^{ N_{\tau_j}}\log (d_i)\right)
\end{aligned}
\end{equation}

Substituting the expressions (\ref{eq7}), (\ref{eq17}), (\ref{eq19}) in the objective function of the expression (\ref{eq9}) we have that,

\begin{align}
	\label{eq20}
	\ln P_{\tau}(\theta) - \ln f_\tau(D\mid\theta) - \ln f_\tau(\theta)
	&= 2\sum_{i=1}^{J+1}\dfrac{\ln(\tau_i-\tau_{i-1})}{2}+  \ln(J) + \sum_{i=2}^J\ln(\tau_i)\nonumber\\
	&-\sum_{j=1}^{J+1} \left(\dfrac{\tau_{j-1}^{\alpha_j}-\tau_{j}^{\alpha_j}}{\beta_j^{\alpha_j}} \right.\nonumber\\
	&\left.+ (N_{\tau_j}-N_{\tau_{j-1}})\left(\ln(\alpha_j)- \alpha_j\ln(\beta_j)\right) \right.\nonumber\\
	&\left.+ (\alpha_j-1)\sum_{i= N_{\tau_{j-1}}+1}^{ N_{\tau_j}}\ln (d_i)\right)\nonumber\\
	&-\sum_{j=1}^{J+1} \left((\phi_{12}-1)\ln\alpha_j - \phi_{11}\alpha_j\nonumber \right.\\
	&\left. + (\phi_{22}-1)\ln\beta_j - \phi_{21}\beta_j\right) + J \ ln (T-1)
\end{align}

Each of the members of the same generation will have a Bayesian-MDL, of which the smallest is chosen. This is done for all the generations. At the end, we will have as many Bayesian-MDLs as generations, and the minimum corresponding to the solution sought in the problem of determining the points of change of the time series of intervals is chosen. Now, we will proceed to address the performance of the algorithm under different escenarios, three simulated and one of real data.

\section{Results and discussion}\label{sec4}

In this section, we proceed to assess the performance of the algorithm to detect multiple change-points. For this purpose we consider two datasets, the first one being simulated observations and the second one consistent of records of particulate matter of less than 2.5 microns of diameter ($PM_{2.5}$) in the city of Bogotá, Colombia collected during the period 2018-2020 on a daily basis.

For the implemented experiments, only the Weibull mean cumulative function (\ref{eq20}) was considered with optimal values for the parameters $\alpha$ and $\beta$, 0.1 and 0.5 respectively, which were estimated via the \texttt{optim} function of the statistical software \texttt{R} \cite{RcoreTeam2022}; on the other hand, for $\alpha$ and $\beta$ we used as prior distributions $Gamma(\phi_{i1}, \phi_{i2}), \ i = 1, 2$ as we previosly defined in section (\ref{Weibull}) and the optimal values for the hyperparameters were found to be  $\phi_{11} = 1$, $\phi_{12} = 2$, $\phi_{21} = 3$ and $\phi_{22} = 1.2$ estimated using Markov-Chain Monte Carlo (MCMC) methods such that, the objective function takes the form of (\ref{eq19}).

We start by analyzing the simulated data under three settings which will be described soon in a detailed manner and then we proceed with the real data. For this task we used the statistical software \texttt{R}; the scripts and datasets can be shared on request.

\subsection{Simulation study}

To assess the performance of the algorithm on simulated data, we proceed under a similar scheme as the one proposed by \cite{Li2012}. Three different settings were considered for the number of change-points $J$, 1, 2 and 3 and their locations, $\tau_1, \tau_2, \ldots, \tau_J$ were selected in a convenient manner to illustrate, such that these are presented in table (\ref{table:SimulationsTableJ}).\

Taking into account that the length of the $PM_{2.5}$ series for Bogotá for 2018-2020, was 1096, such number of observations were simulated from a \textit{log-normal} distribution with scale parameter $\mu \in \mathbb{R}$ and of shape $\sigma > 0$ (see expression (\ref{eq:log-normal})). The data was approximated to this distribution as per the results obtained using the library \texttt{fitdistrplus} \cite{Delignette2015} in \texttt{R} and the ones in \cite{Rumburg2001}.

\begin{equation}\label{eq:log-normal}
  f(x) =
  \begin{cases}
    \frac{1}{x \sigma \sqrt{2 \pi}} exp \Bigg( -\frac{(ln(x) - \mu)^2}{2 \sigma^2} \Bigg), & \text{$x > 0$}\\
    0, & x < 0
  \end{cases}
\end{equation}

\begin{figure}[ht]
\centering
\includegraphics[width=0.8\textwidth]{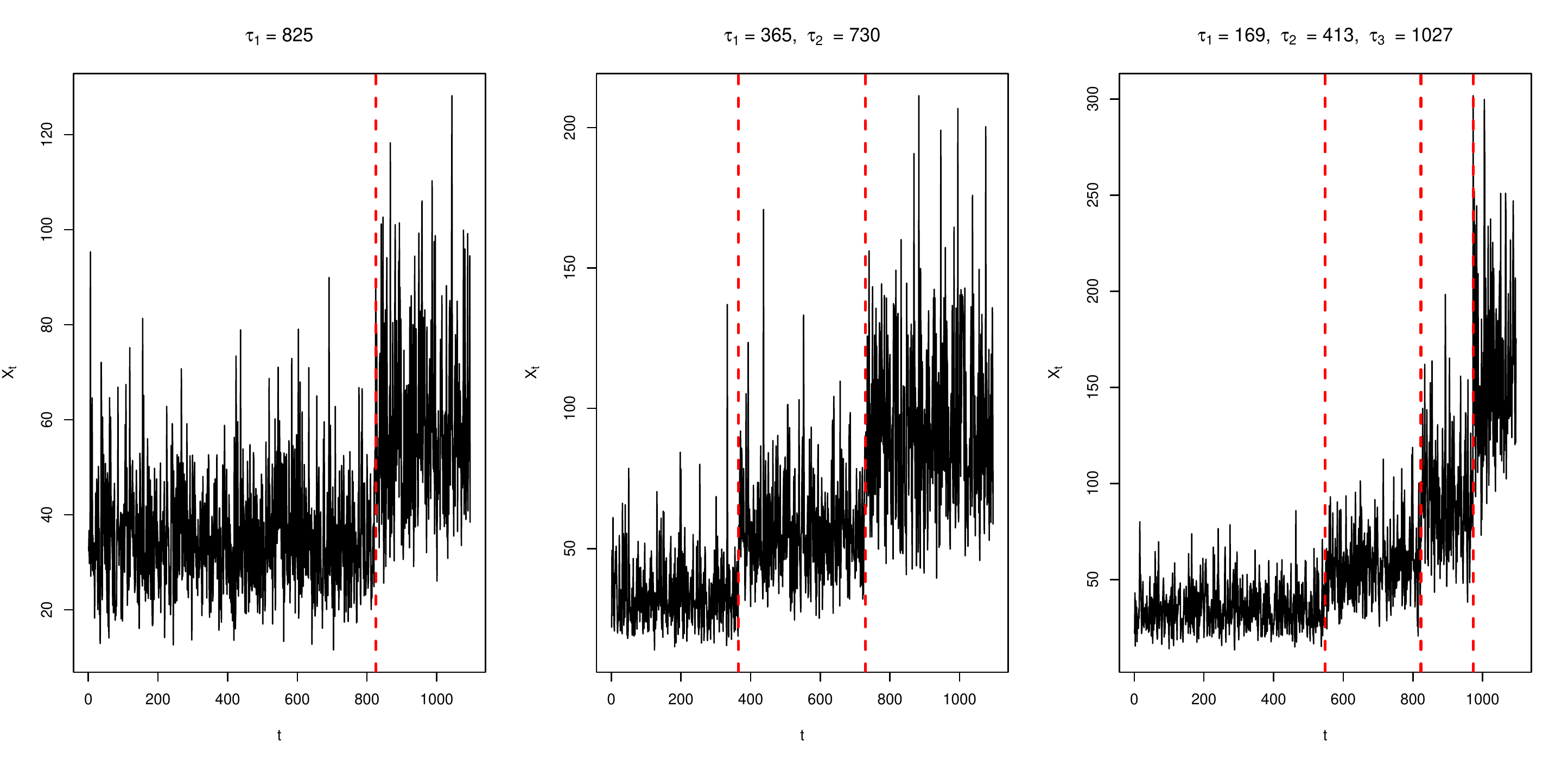}
\caption{Simulated time series behavior}
\label{fig:CPSimulationSettings}
\end{figure}

Now, let's remember that $J$ change-points split the time series into $J+1$ sub-series or regimes, such that for each $J$, every regime was generated by incrementally varying the scale parameter $\mu$ in 0.5 units while the parameter $\sigma$ was held constant and equal to 0.32. Thus, the values of $\mu$ and $\sigma$ used to generate the $J+1$ regimes, $J \in \{1, 2, 3\}$ are presented in table (\ref{table:SeriesParameters}) while the series graphical behavior for the three settings of $\tau_1, \tau_2, \ldots, \tau_J$ can be appreciated in figure (\ref{fig:CPSimulationSettings}) such that the vertical dashed lines represent the change-points. On the other hand, again to illustrate, we defined as threshold for possible exceedances the arithmetic mean of the 1096 simulations, $\bar{X} = \frac{1}{1096} \sum_{t = 1}^{1096} X_t$.

For each of the three settings, the number of change-points were estimated, the optimal Bayesian-MDL and the cumulative mean function, $m(t\mid\theta)$ and for every run of the genetic algorithm, 50 generations with 50 individuals each were used; the proportion of the times used to generate the initial population was of 6\% and the mutation probability was 3\% as in \cite{Li2012}. These results are presented and analyzed in  the following section.

\begin{table}[h]
\begin{center}
\begin{tabular}{cc}
\hline
\textbf{Number of Change-points} & \textbf{Locations}  \\ \hline
1                          & $\tau_1 = 825$ \\
\hline
2                         & $\tau_1 = 365$, $\tau_2 = 730$  \\
\hline
3                        & $\tau_1 = 548$, $\tau_2 = 823$, $\tau_3 = 973$  \\
\hline                
\end{tabular}
\end{center}
\caption{Different Change-points simulations considered}
\label{table:SimulationsTableJ}
\end{table}

\begin{table}[h]
\begin{center}
\begin{tabular}{p{3.9cm}p{3cm}p{4.2cm}}
\hline
\textbf{Number of Change-points} & \textbf{Number of regimes} & \textbf{Regime distribution}   \\ \hline
1               & 2 &  $log-normal(\mu = 3.5, \sigma = 0.32)$, $log-normal(\mu = 4.0, \sigma = 0.32)$ \\
\hline
2               & 3  & $log-normal(\mu = 3.5, \sigma = 0.32)$, $log-normal(\mu = 4.0, \sigma = 0.32)$, $log-normal(\mu = 4.5, \sigma = 0.32)$ \\
\hline
3               & 4  & $log-normal(\mu = 3.5, \sigma = 0.32)$, $log-normal(\mu = 4.0, \sigma = 0.32)$, $log-normal(\mu = 4.5, \sigma = 0.32)$, $log-normal(\mu = 5.0, \sigma = 0.32)$ \\
\hline
\end{tabular}
\end{center}
\caption{Settings for the time series regimes for different number of change-points}
\label{table:SeriesParameters}
\end{table}

The results of the implementation of the proposed model in the three different schemes with simulated data and the real data set in the context of air quality will be shown below. For this, the performance on each data set will be illustrated by means of figures (\ref{fig:Simulation1CP3}), (\ref{fig:Simulation2CP3}), (\ref{fig:Simulation3CP2}) and (\ref{fig:fig2}). These figures should be interpreted as follows. The graph on the upper left represents the behavior of the observed cumulative mean number of overshoots (in black), compared to the estimated cumulative mean function $m(t\mid\hat{\theta})$ (in red) and their respective confidence intervals of the estimated at $95\%$ (in blue). At the top right of the same figure is a plot representing the behavior of BMDL across each of the 50 generations. In the lower left is the histogram that determines the number of times a change point is repeated over the 50 generations. The lower right graph relates the change-points obtained in each of the 50 generations. It is important to note that the simulations shown in the figure (\ref{fig:CPSimulationSettings}) are intended to show three of the most important schemes.  In the first panel with only one change point, we want to show the case in which there is a significant disproportion between the observations to the left of the change point, versus those to the right of the change point ($\tau_{1}=825$). In the second scheme with the two change-points $\tau_1 = 365$, $\tau_2 = 730$ we want to exhibit the case where the points evenly split the segments in the time series. While for the last scheme it is sought to establish the predictive ability of the algorithm in the case where the change-points are at the end of the time series as is the case of the scheme where the location of the same are in $\tau_1 = 548$, $\tau_2 = 823$, $\tau_3 = 973$. In the later scheme it should be noted that all the change-points are after the middle of the initial observations of the time series of interest.

\subsubsection{First Setting: One Change-point}

\begin{figure}[ht]
\centering
\includegraphics[width=0.8\textwidth]{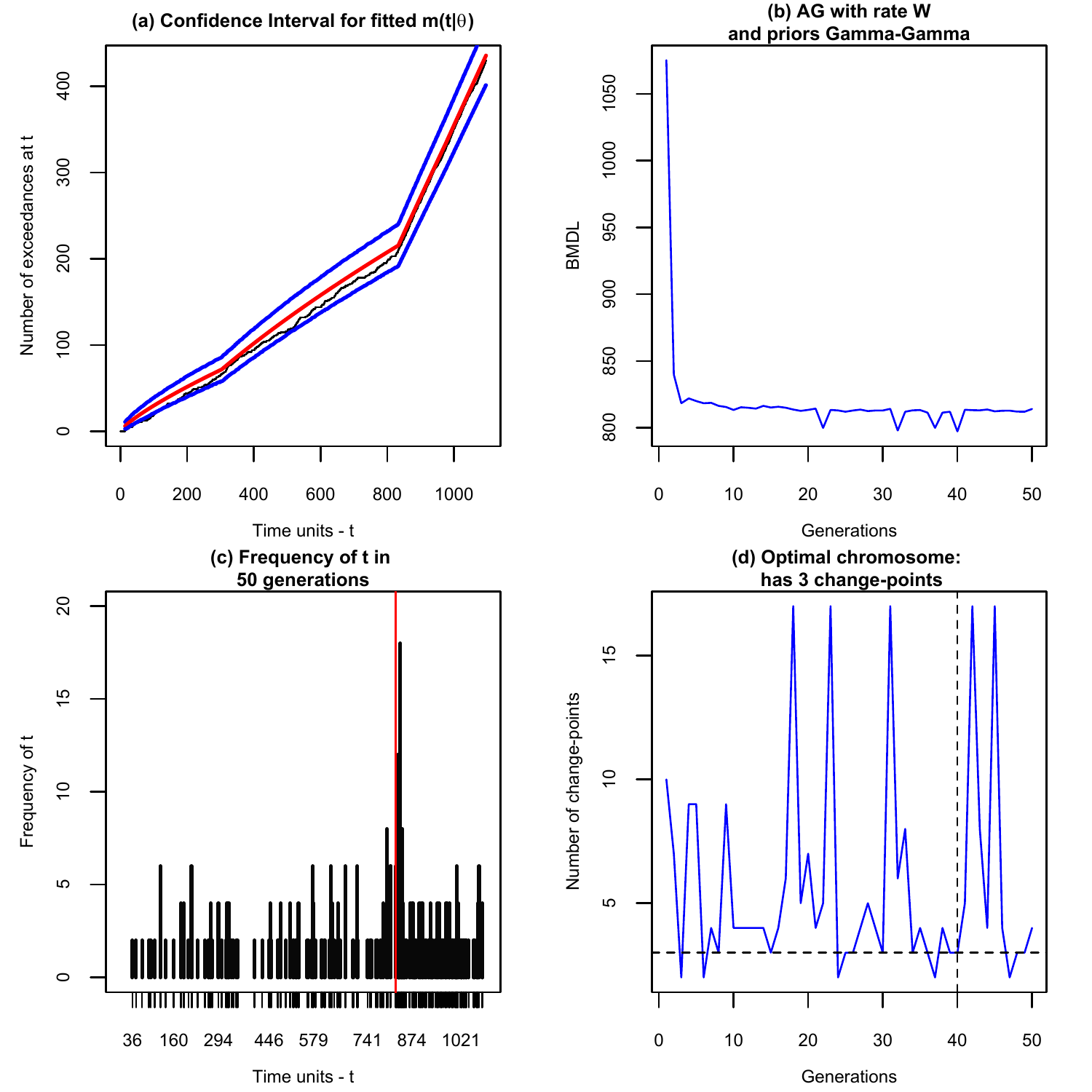}
\caption{Genetic algorithm results - 1 change-points: {\bf (a)} Confidence interval for fitted $m(t\mid\theta)$ (blue lines), estimated cumulative average value of the number of exceedances at time $t$ (red line) and number of exceedances observed at time $t$ (black line). {\bf (b)} BMDL calculated for each of 50 generations. {\bf (c)} Frequency of the number of times each of the change-points is observed in the 50 generations. The red lines indicate the real locations of the change-points. {\bf (d)} Estimated number of change-points in each of the 50 generations.The dashed lines indicate the optimal estimated number of change-points.}

\label{fig:Simulation1CP3}
\end{figure}

\begin{figure}[ht]
\centering
\includegraphics[width=0.8\textwidth]{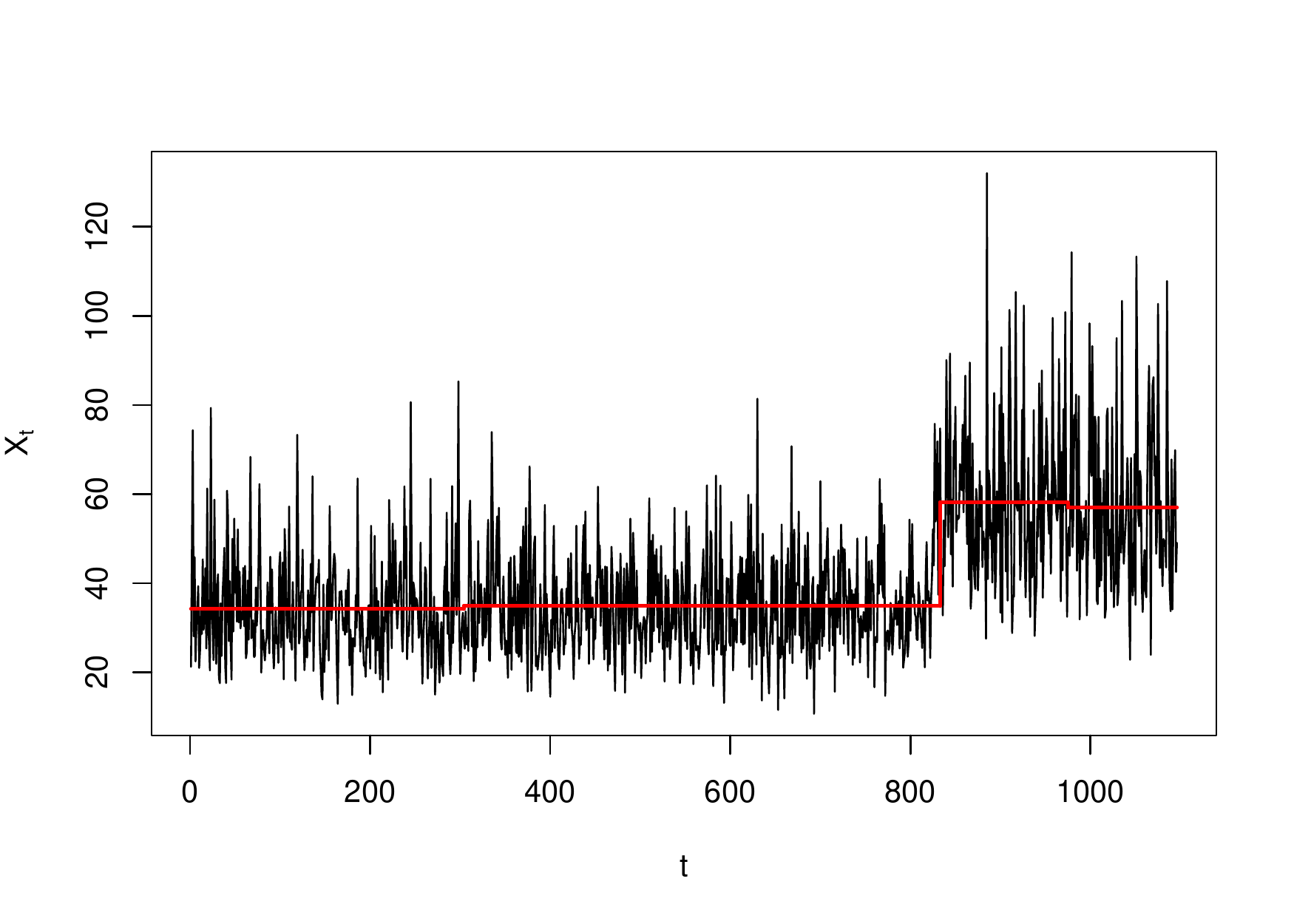}
\caption{First setting - Mean fitting for the regimes}
\label{fig:Simulation1CP3Breaks}
\end{figure}

We start by considering the first configuration of the table \ref{table:SimulationsTableJ}, i.e. $\tau_1 = 825$. In the figure (\ref{fig:Simulation1CP3}) the results of the genetic algorithm implementation are recorded. The top left plot shows the good fit of the observed cumulative overruns with the estimated $m(t\mid\hat{\theta})$ function, before and after $\tau_1 = 825$. The estimated value of the number of overshoots at $\tau_1 = 825$ is $213.51$ (in red) versus the observed value of $204$ (in black). All observed values are contained within the $95\%$ confidence bands (in blue).

On the other hand, according to the next plot (upper right panel) the minimum reached for the Bayesian-MDL was found around the 40th generation taking a value of 797.27; such generation is marked explicitly in the lower right panel by the vertical dashed line while the optimal number of change-points detected was $J = 3$ as per the horizontal dashed line.

Despite the real value  of $J = 1$ being different to the estimated one, in the last panel (lower left) the bar associated with the more  frequent change-point can be found in a neighborhood of $\tau_1 = 825$, furthermore, taking as a reference such real value (vertical solid line) we have that the estimations are in average, two units over the optimum. In the same histogram, it can be seen that of the 50 repetitions (one for each generation), 20 are very close to $\tau_1 = 825$.

Now, the figure (\ref{fig:Simulation1CP3Breaks}) shows using the solid horizontal line, the sample mean associated to each regime defined by the estimated change-points. Thus the algorithm accomplishes for the whole series capturing the trend of the values under the presence of said points even while there are differences regarding the estimated $J$ and its real value.

\subsubsection{Second Setting: Two Change-points}

\begin{figure}[ht]
\centering
\includegraphics[width=0.8\textwidth]{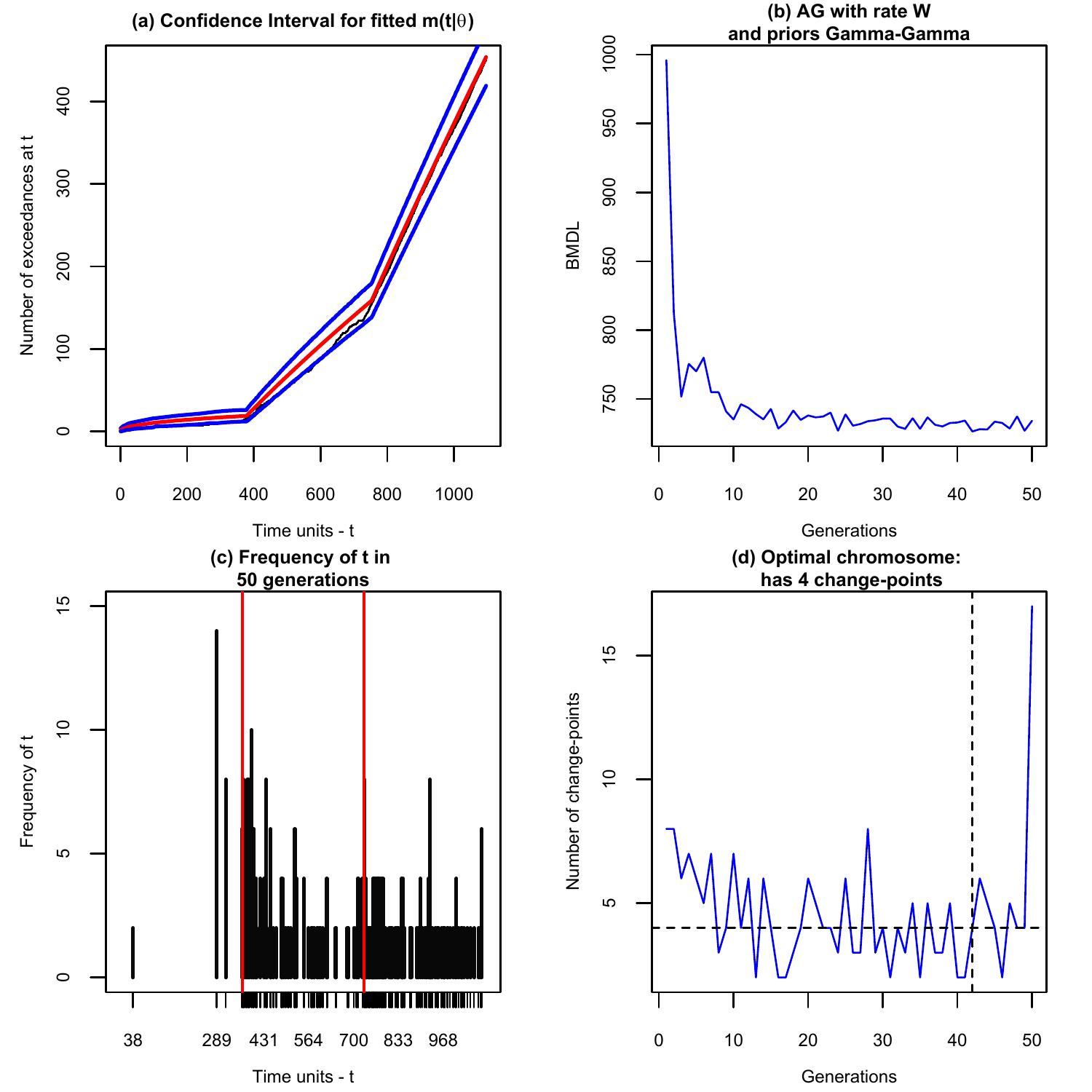}
\caption{Genetic algorithm results - 2 change-points: {\bf (a)} Confidence interval for fitted $m(t\mid\theta)$ (blue lines), estimated cumulative average value of the number of exceedances at time $t$ (red line) and number of exceedances observed at time $t$ (black line). {\bf (b)} BMDL calculated for each of 50 generations. {\bf (c)} Frequency of the number of times each of the change points is observed in the 50 generations. The red lines indicate the real locations of the change points. {\bf (d)} Estimated number of change points in each of the 50 generations.The dashed lines indicate the optimal estimated number of change points.}
\label{fig:Simulation2CP3}
\end{figure}

\begin{figure}[ht]
\centering
\includegraphics[width=0.8\textwidth]{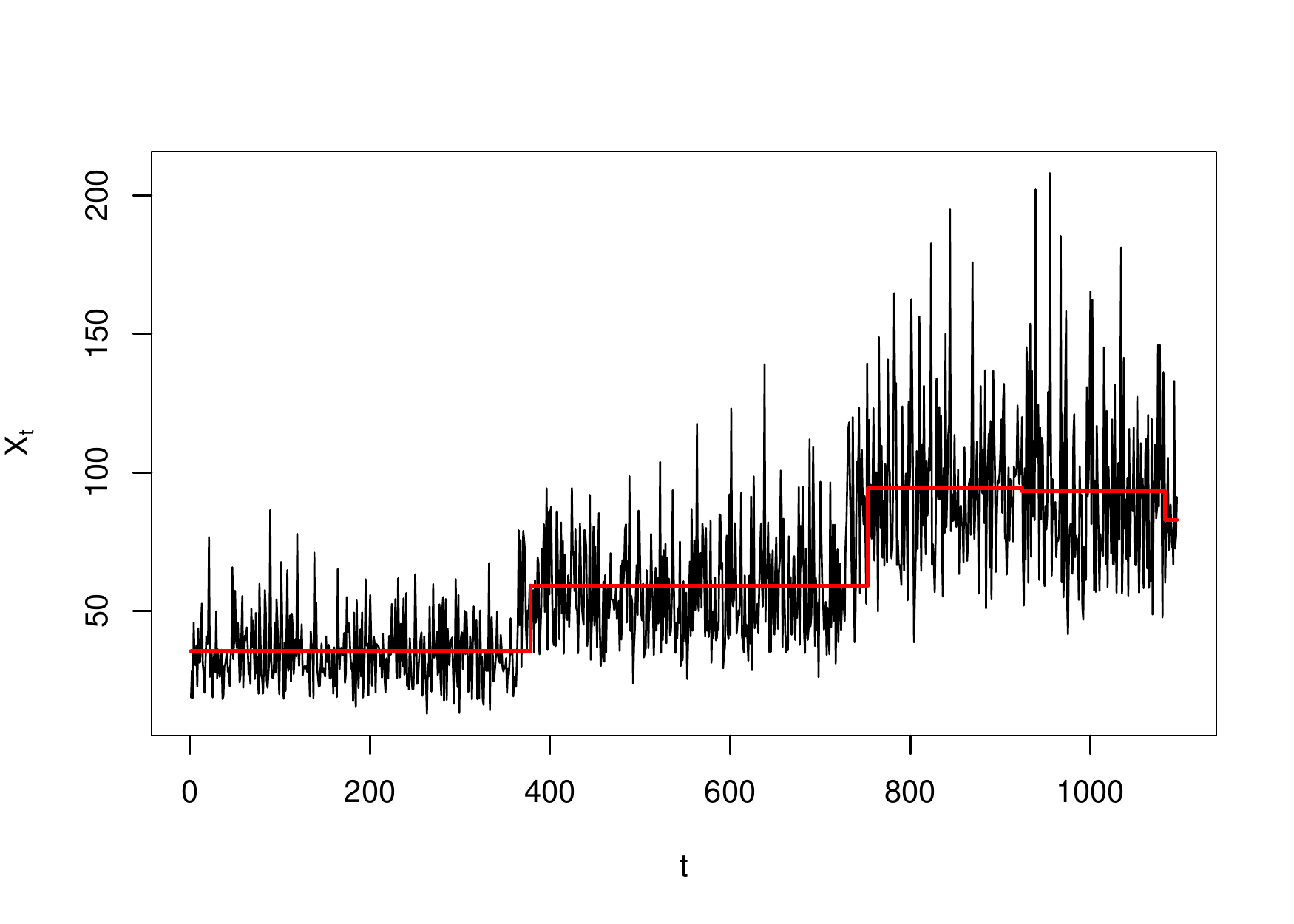}
\caption{Second setting - Mean fitting for the regimes}
\label{fig:Simulation2CP3Breaks}
\end{figure}

Now we have the second setting of the table (\ref{table:SimulationsTableJ}), $\tau_1 = 365$ and $\tau_2 = 730$ with the results of the implementation of the  genetic algorithm presented in the figure (\ref{fig:Simulation2CP3}). We start by analyzing the fitting of the mean cumulative function (upper left panel) and move in a clockwise manner.

In item (a) of the figure (\ref{fig:Simulation2CP3}) there is a good fit of  $m(t\mid\theta)$ (red line) so that its values overlap with the real ones (black line), while the $95\%$ confidence bands (blue lines) completely contain the real observations. Again, as in the previous case, these estimates grow in a stepwise fashion following the behavior of the actual values with breaks defined by the change points. The estimate of $m(t\mid\theta)$ over the change points were $m(\tau_1\mid\hat{\theta}) = 18.54$, $m(\tau_2\mid\hat{\theta}) = 150.63$. That is, we estimate about 19 overshoots up to $\tau_1$ and about 151 overshoots up to $\tau_2$, compared to the observed values of $12$ in $\tau_{1}$  and $137$ in $\tau_{2}$.


The following graph (b) of (\ref{fig:Simulation2CP3}), presents the evolution with respect to the generations of the Bayesian-MDL, reaching the minimum in generation 42 of the 50. It follows from there to observe in d. of the same figure, that the chromosome of generation 42, suggests $J = 4$ change points, when in reality they are only 2 change points. However, as can be seen in histogram c. of the same figure, it is noted that the highest frequencies in the 50 generations contain very small neighborhoods of the true values of $\tau_{1}$ and $\tau_{2}$.

%
%
On the other hand, (\ref{fig:Simulation2CP3Breaks}) through the solid horizontal line the sample mean is again represented for each of the regimes after estimating the change-points; as in the previous case the algorithm allows for a proper fitting of the observed data and the variations of the statistic of interest under the presence of said times despite the estimation of $J$ which was superior to the real value.
\subsubsection{Third Setting: Three Change-points}

\begin{figure}[ht]
\centering
\includegraphics[width=0.8\textwidth]{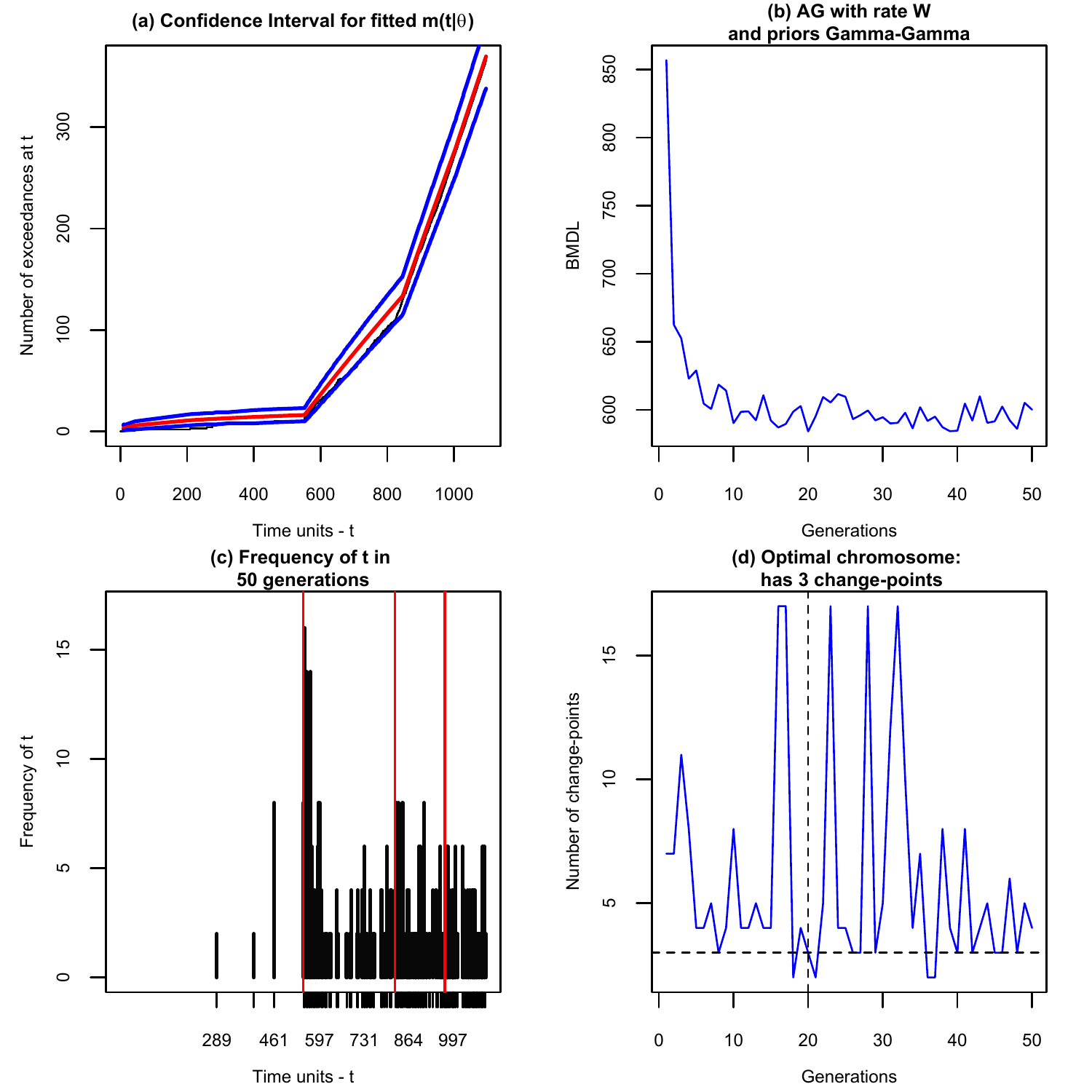}
\caption{Genetic algorithm results - 3 change-points: {\bf (a)} Confidence interval for fitted $m(t\mid\theta)$ (blue lines), estimated cumulative average value of the number of exceedances at time $t$ (red line) and number of exceedances observed at time $t$ (black line). {\bf (b)} BMDL calculated for each of 50 generations. {\bf (c)} Frequency of the number of times each of the change points is observed in the 50 generations. The red lines indicate the real locations of the change points. {\bf (d)} Estimated number of change points in each of the 50 generations.The dashed lines indicate the optimal estimated number of change points.}

\label{fig:Simulation3CP2}
\end{figure}

\begin{figure}[ht]
\centering
\includegraphics[width=0.8\textwidth]{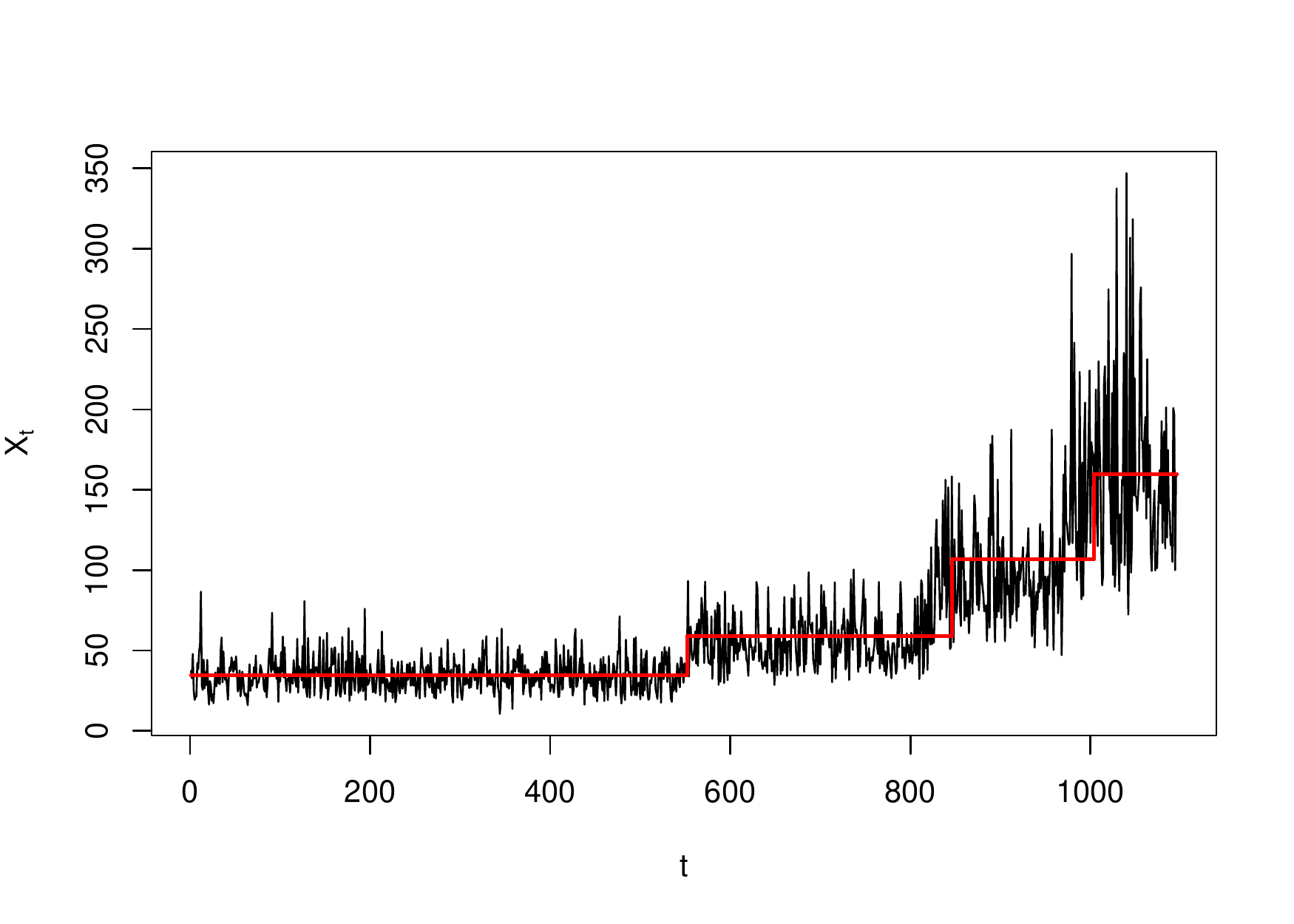}
\caption{Third setting - Mean fitting for the regimes}
\label{fig:Simulation3CP2Breaks}
\end{figure}

Finally we have now the case of three change-points such they are located at $\tau_1 = 169, \tau_2 = 413$ and $\tau_3 = 1027$ and the results of applying the genetic algorithm are presented in figure (\ref{fig:Simulation3CP2}).
In this case, it is worth noting that as seen in the time series depicted in the figure (\ref{fig:Simulation3CP2Breaks}), the three simulated changes are presented back-to-back at the end of the series, making it more difficult to capture the actual location of the change points $\tau_1 = 169, \tau_2 = 413$ and $\tau_3 = 1027$. This apparent difficulty, can be seen reflected in the fit of the function  $m(t\mid\theta)$, in panel a. of the figure (\ref{fig:Simulation3CP2}), since it does not appear that what is observed is fully contained within the $95\%$ confidence bands, at the beginning of the series. Estimated values were $m(\tau_1\mid\hat{\theta}) = 15.12$, $m(\tau_2\mid\hat{\theta}) = 123.76$, $m(\tau_1\mid\hat{\theta}) = 250.67$, versus the actual observed $10$, $109$ and $244$ respectively, showing high predictive ability of the model, even in the presence of many changes at the end of the series of interest.


Now we analyze the evolution of the Bayesian-MDL for each generation of the genetic algorithm (upper right panel) its behavior seems to be approximately constant for all its path with some decreases such that the optimal value was 584.13. Said optimum was reached in generation 20 as noted by the vertical dashed line in the third plot (lower right panel); on the other hand, the optimal number of change-points was found to be $J = 3$ according to the horizontal dashed line in the panel d., the real number of change-points.

Similarly as in the other two cases, the estimated change-points cluster around the real ones as per the last plot of (\ref{fig:Simulation3CP2}) (lower left panel) and as per the values around the vertical solid lines representing the real optimum; as in the other cases the more frequent values are far from the real optimum in average in just two units or one, this for the first two change-points while for the third one the algorithm captures such value with difficulty which is consequent with the behavior for the mean cumulative function. Finally, in (\ref{fig:Simulation3CP2Breaks}) we have the fitting of the sample mean for all the considered regimes; the trend associated to the change-points it is not only captured by the algorithm but also it can be appreciated how the found ruptures are in a neighborhood close to the real change-points which is consistent with the results in the last plot of (\ref{fig:Simulation3CP2}).

Additionally to the three presented settings, in the appendix in the third section (\ref{sec:NumberOfChangePointsExperiment}), the reader can found some results in regards of the performance of the algorithm with a larger number of change-points, i.e., $J = 10, 20, \ \text{and} \ 50$ as the structure of the objective function (\ref{eq20}) depends on it. Further experiments present in the apendix in the first section (\ref{sec:ComparisonExperiments}) were also conducted with other available methods such as the Pruned Linear Exact Time (PELT) \cite{KillickEtAl2012}, the Cumulative Sums (CUSUM) \cite{Page1954} and the original that this work is based on by \cite{Li2012} showing the here proposed approach is robust to deviations from distributional assumptions

\subsection{Real Data Analysis}

\begin{figure}[h] 
\begin{center}
\includegraphics[width=0.8\textwidth]{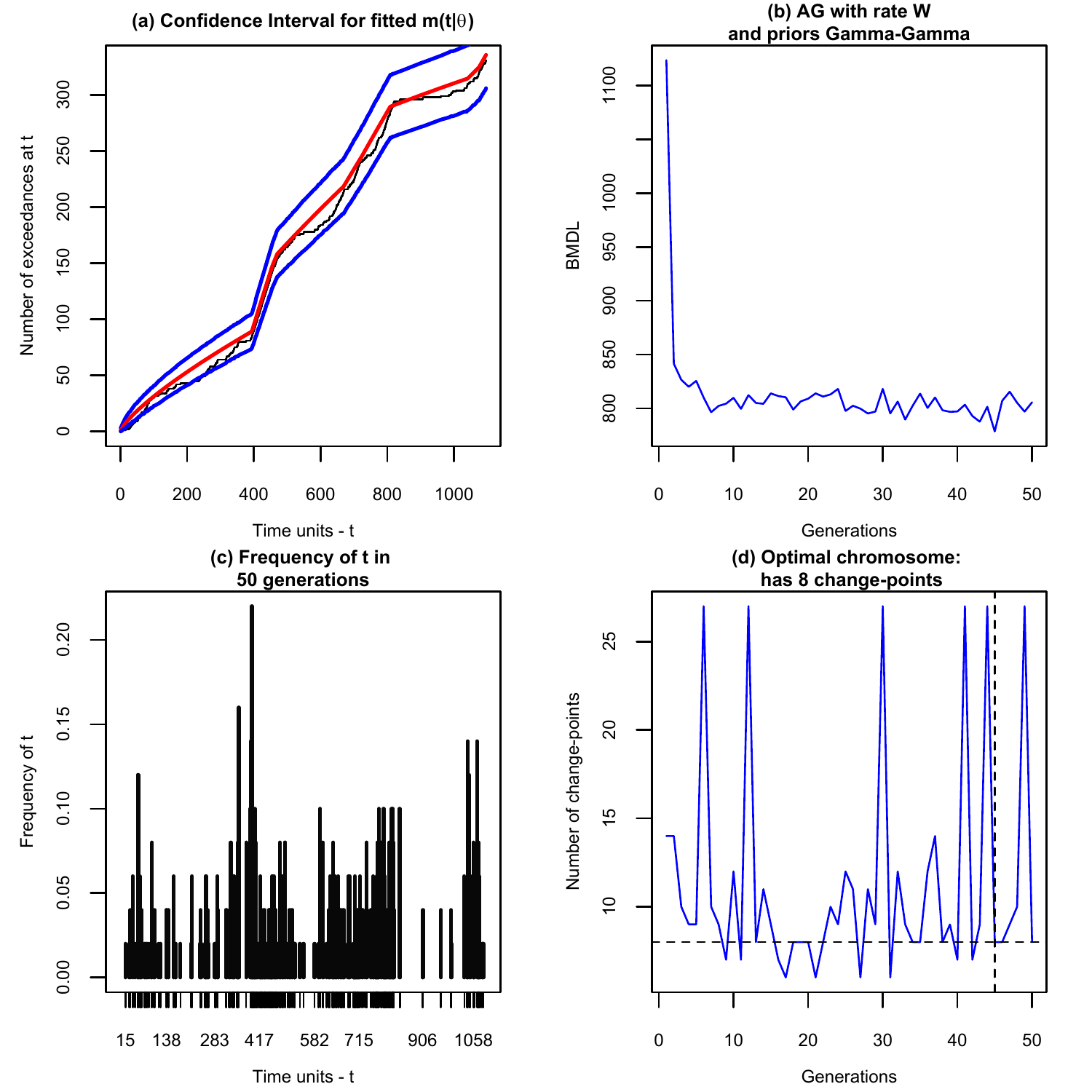}
\caption{$PM_{2.5}$ Measurements in Bogota, January 1 2018 - December 31 2020: {\bf (a)} Confidence interval for fitted $m(t\mid\theta)$ (blue lines), estimated cumulative average value of the number of exceedances at time $t$ (red line) and number of exceedances observed at time $t$ (black line). {\bf (b)} BMDL calculated for each of 50 generations. {\bf (c)} Frequency of the number of times each of the change points is observed in the 50 generations. {\bf (d)} Estimated number of change points in each of the 50 generations. The dashed lines indicate the optimal estimated number of change points.}
\label{fig:Daily}
\end{center}
\end{figure}

Here we applied on the $PM_{2.5}$ series for Bogotá in the period 2018 - 2020, in an exhaustive manner, the genetic algorithm with the already described settings at the beginning of this section and using as well as an objective function (\ref{eq19}). As threshold, the one established by the environmental norm for Colombia was used which specifies that presence of this polluting agent can not overpass  $37 \mu g / m^3$. The series can be observed in figure (\ref{fig:Daily}).

Then, the optimal chromosome was found to be (8, 400, 408, 445, 488, 627, 654, 661, 798), such that there are $J = 8$ change-points. The horizontal solid line in the third plot (lower right panel) of figure (\ref{fig:Daily}) indicates the sample mean of every regime such that the start of each one of these represents the presence of a change-point, except for the first one and the last times. It can be observed as well, that the first 4 change-points are very close and they overlap. The table (\ref{table:Points}) shows the date corresponding to each change-point.

\begin{table}[h]
\begin{center}
\begin{tabular}{ccc}
\hline
\textbf{Change Point} & \textbf{Day of the week} & \textbf{Date}      \\ \hline
400                   & Monday                   & February 4, 2019   \\
408                   & Tuesday                  & February 12, 2019  \\
445                   & Thursday                 & March 21, 2019     \\
488                   & Friday                   & May 3, 2019        \\
627                   & Thursday                 & September 19, 2019 \\
654                   & Wednesday                & October 16, 2019   \\
661                   & Wednesday                & October 23, 2019   \\
798                   & Sunday                   & March 8, 2020      \\
\hline
\end{tabular}
\end{center}
\caption{Points of change with their respective dates}
\label{table:Points}
\end{table}

These surpasses occurred during the rainy season in Bogotá. If the dates in the table above are compared with the measurements in figure (\ref{fig:Daily}), the seasons with the most consistent peaks over time can be captured.

\begin{figure}[h]
\begin{center}
\includegraphics[width=0.8\textwidth]{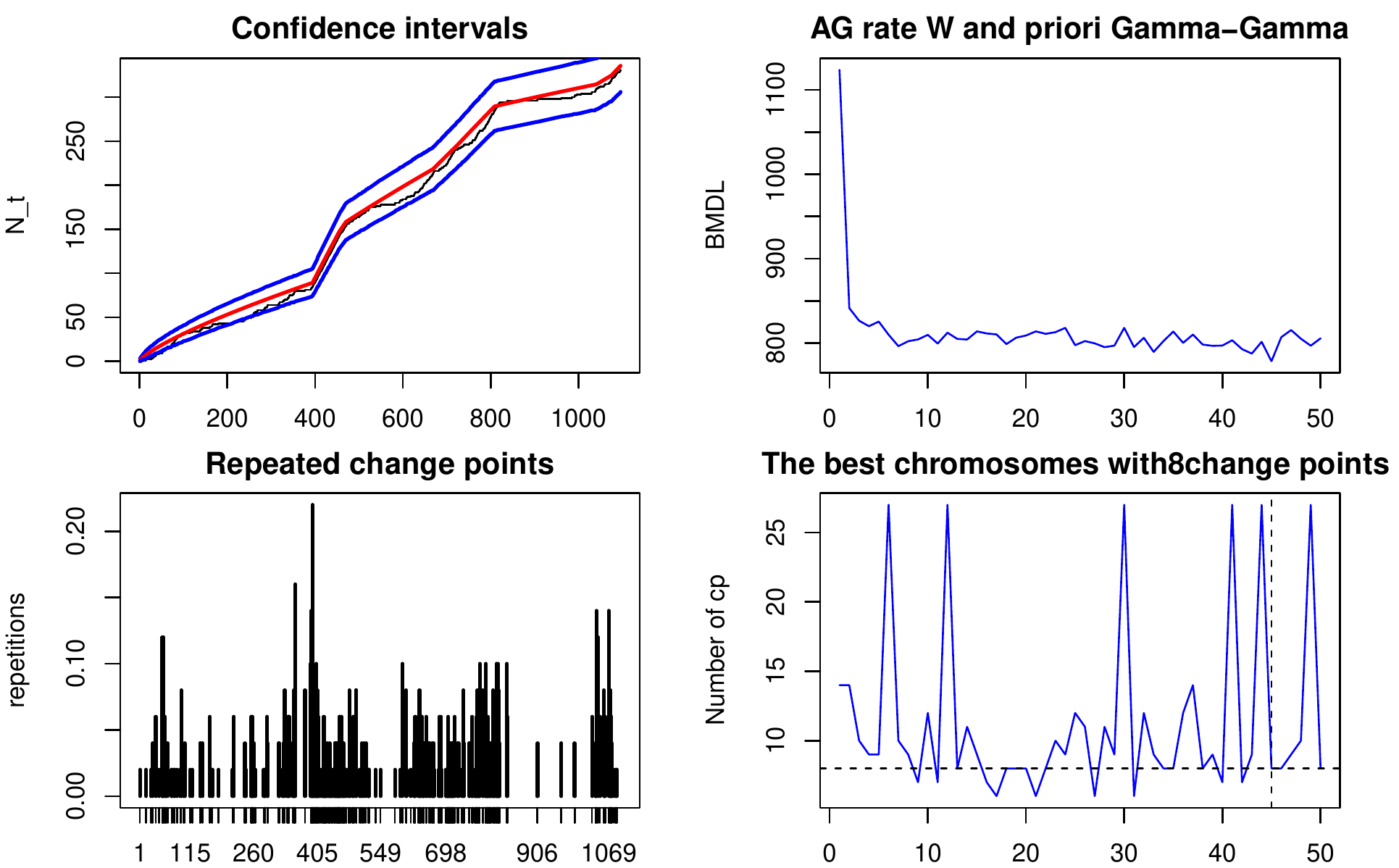}
\end{center}
\caption{Actual data on $PM_{2.5}$ pollution in Bogota from 2018 to 2020. {\bf (a)} Confidence interval for fitted $m(t\mid\theta)$ (blue lines), estimated cumulative average value of the number of exceedances at time $t$ (red line) and number of exceedances observed at time $t$ (black line). {\bf (b)} BMDL calculated for each of 50 generations. {\bf (c)} Frequency of the number of times each of the change points is observed in the 50 generations. The red lines indicate the real locations of the change points. {\bf (d)} Estimated number of change points in each of the 50 generations.The dashed lines indicate the optimal estimated number of change points.}

\label{fig:fig2}
\end{figure}

As can be seen in the other plots in (\ref{fig:fig2}), the one in the upper left panel shows the adjustment of the cumulative mean function $m(t\mid\theta)$ (black line) by means of the point averages, obtained from the vectors of parameters better qualified by the MDL (red line), and their 95\% confidence interval (blue lines).

Also, the corresponding optimal Bayesian-MDL was 778.44 as per the upper right panel which shows the evaluation of said metric for each one of the best chromosomes of each of the 50 generations such that the minimum is reached around the 45th one (vertical dashed line in the lower right panel).

Finally, in the lower left panel, the histogram shows the days in the time series that exceed the threshold, which appear most frequently in the first 50 chromosomes of the 50 generations analyzed.

\begin{figure}[h]
\begin{center}
\includegraphics[width=0.7\textwidth]{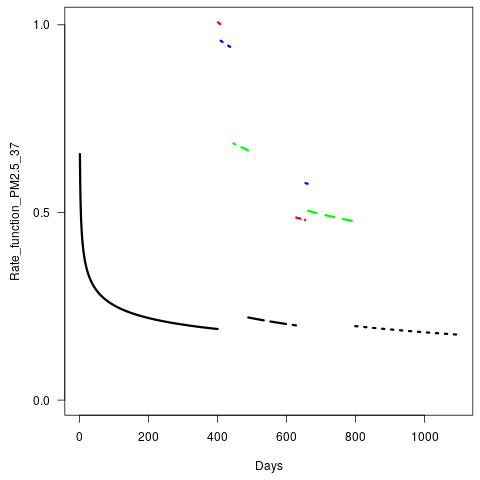}
\end{center}
\caption{Rate function for $PM_{2.5}$ by Days}
\label{fig:fig3}
\end{figure}

The graph in figure (\ref{fig:fig3}) shows that before day 400, i.e., Monday, February 4, the rate of exceedances of the 37 $\mu$g/$m^3$ threshold had been decreasing sharply, but after it, the highest emission rate for eight consecutive days was recorded. This high average rate is around 1.004, as shown in (Table\ref{table:Regime}) for the second regime. In the third regime, it decreased to an average rate of 0.9468; in the fourth, it jumps to 0.6748, and it is in the fifth regime that it achieves the lowest drop, 0.2090, before the intensity function rises again to levels above the 0.2 threshold overshoots per unit of time. This  rate present in the fifth regime goes from May 3 to September 19, 2019. Thus, it is evident that the fifth regime occurred before the COVID-19 pandemic lockdown was declared in Colombia. Therefore, this may be the result of a public policy aimed at reducing the emissions of $PM_{2.5}$.

As part of the 2010-2020 ten-year plan for air pollution control, the use of emission control systems in cargo transport vehicles and motorcycles, and as well as the  integrated public transport system (SITP for its acronym in Spanish) policies were implemented. The later includes the replacement of old buses with internal combustion engine with electric or hybrid buses. In addition to the above, a few days before the fifth regime, resolution 383 (see RES19 (2019)) was issued, which declared a yellow alert for particulate ma\-tter exceedances. Considering the first regime represented in figure (\ref{fig:fig3}), the rapid deceleration in the emission of threshold exceedances can also be seen as a consequence of this resolution. Such is the case of restrictions on the use of transportation and the mobility sector, in addition to those aimed at the operations of industries that use combustion processes associated mainly with the burning of biomass and the use of fossil or liquid fuels.

\begin{table}[h]
\begin{center}
\begin{tabular}{cccc}
\hline
\textbf{Regime} & \textbf{Min} & \textbf{Mean} & \textbf{Max} \\ \hline
1               & 0.1894       & 0.2379        & 0.6555       \\
2               & 1.002        & 1.004         & 1.007        \\
3               & 0.9363       & 0.9468        & 0.9577       \\
4               & 0.6657       & 0.6748        & 0.6841       \\
5               & 0.1991       & 0.2090        & 0.2201       \\
6               & 0.4801       & 0.4832        & 0.4864       \\
7               & 0.5765       & 0.5773        & 0.5781       \\
8               & 0.4761       & 0.4896        & 0.5043       \\
9               & 0.1744       & 0.1849        & 0.1971       \\
\hline
\end{tabular}
\end{center}
\caption{Minimum, Maximum and Mean for each Regime}
\label{table:Regime}
\end{table}

\section{Conclusions and future work}\label{sec5}

A solution has been presented to determine change points in counting time series, particularly when they exceed an environmental standard of interest. So, the need for specifying a likelihood function was overpassed by modeling the times of the series where existed exceedances through a non-homogeneous Poisson process and this specification allowed for better estimations of the number of change-points with a lower variability and a better estimation of parameters of the underlying generating process.

A family of such objective functions was derived using different definitions for the mean cumulative function of the non-homogeneous Poisson processes but it is yet to be seen  the influence of using one or the other on the estimation of the change-points and their location.

The search method for the change-points was a genetic algorithm similar to the one exposed by \cite{Li2012}, nevertheless, there is still room for improvement for it in terms of different operators available at literature and as well the use of other families of such algorithms.

The detection of change-points using a genetic algorithm and a Bayesian-MDL as selection criterion yielded results that agree with the public policies implemented in Bogotá, Colombia, both regarding the contamination alerts issued by the monitoring network, as well as the mobility restrictions due to the quarantine caused by the SARS-CoV-2 virus pandemic.

The good performance of the algorithm is partly due to the fact that in \cite{Achcar10} and \cite{Achcar11}, the cumulative means function of the contamination exceedances was adjusted, but by then there was no computational technique for the automatic detection of change-points.

We hope that in the near future these methods will prove to be a useful tool for the government agencies in charge of measuring the effectiveness of the actions taken to reduce air pollution, and thus reduce its impacts both in the environment and the health of the inhabitants.

\section*{Conflict of interest}
The authors declare that they have no conflict of interest.

\bibliographystyle{unsrt}
\bibliography{bibliography}

\newpage

\appendix 

\section{Appendix}

\subsection{Comparison with Other Methods}
\label{sec:ComparisonExperiments}

In this section, we compare the performance of the genetic algorithm proposed in the paper with three previously documented methods, two widely used in the literature, the Pruned Exact Linear Time (PELT) \cite{KillickEtAl2012} and the Cumulative Sum (CUSUM) \cite{Page1954} and the genetic algorithm originally proposed by \cite{Li2012}. For this purpose we again use the three data sets simulated in section (4) of the paper that correspond to realizations of $log-normal(\mu, \sigma)$ distributions.

\subsubsection{Description of the Methods}

\textbf{Pruned Exact Linear Time (PELT) \cite{KillickEtAl2012}} \\

Consider again a time series, $y = y_t = \{y_1, y_2, \ldots, y_T \}$.. This will have $J$ change-points with positions, $\mathcal{T} = \{\tau_1, \tau_2, \ldots, \tau_J \}, \tau_1 < \tau_2, \ldots, \tau_J$. In the presence of such instants, the series will be divided into $J+1$ subseries or regimes, whose observations are bounded by $\tau_j, \tau_{j+1}, \ j  = 0, 1, 2, \ldots, J$  with  $\tau_0$  the time one and  $\tau_{J+1}$ the observation corresponding to time $T$. Then, it is of interest to optimize a function with the following structure,	
	\begin{equation}
	\sum_{j = 1}^{J+1} [\mathcal{C} (y_{\tau_{j-1}: \tau_j})] + \beta f(J)
	\label{eq:PELT_OF}
	\end{equation}	
	Where $\mathcal{C}(\cdot)$  in (\ref{eq:PELT_OF}) is a cost function. This cost function is commonly taken to be twice the negative log-likelihood of the observations, although there are other definitions including quadratic loss function, cumulative sums function, and log-likelihood of the regimen and regimen length. On the other hand, $\beta f(J)$  corresponds to a penalty and this is chosen in such a way as to guarantee linearity at the change-points, that is, $\beta f(J) = \beta J$ and includes criteria such as the Akaike information criterion (AIC) or the Bayesian Information Criterion (BIC). 	
Now, let $F(s)$ be the model that minimizes (\ref{eq:PELT_OF}) for a data set $y_{1:s}$ and let $\mathcal{T}_s = \{ \tau: 0 = \tau_0 < \tau_1 < \ldots < \tau_J < \tau_{J+1} = s \}$  be the set of possible change-points vectors for such data. Finally, let $F(0) = - \beta$, it then follows that,	
	\begin{equation}
	\begin{split}
	F(s) &= \min_{\mathbf{\mathcal{T}} \in \mathcal{T}_s} \Big \{ \sum_{j = 1}^{J+1} [\mathcal{C} (y_{(\tau_{j-1} + 1):\tau_j} + \beta ] \Big\} \\
	&= \min_t \Bigg \{ \min_{\mathbf{\mathcal{T}} \in \mathcal{T}_t} \sum_{j = 1}^J [\mathcal{C}y_{(\tau_{i-1} + 1):\tau_i}) + \beta]  +  \mathcal{C}(y_{t+1}:n) + \beta \Bigg\} \\
	&= \min_t \{ F(t) + \mathcal{C}(y_{(t+1):T}) + \beta \}
	\end{split}
	\end{equation}

\begin{theorem}
We assume that when introducing a change-point into a sequence of observations the cost, $\mathcal{C}$, of the sequence reduces. More formally, we assume there exists a constant $K$ such that for all $t < s < M$,

\begin{equation}
\mathcal{C}(y_{(t+1):s}) + \mathcal{C} (y_{(s+1):M}) + K \leq C(y_{(t+1):M}) 
\label{theorem:PELTTheorem}
\end{equation}

Then if 

$$F(t)  + \mathcal{C} (y_{(t+1):s}) + K \geq F(s)$$

holds at a future time $M > s$, $t$ can never be the optimal last change-point  prior to $M$.

\end{theorem}

Thus, we have the PELT method whose procedure is explained in the pseudocode (\ref{alg:PELT}). On the other hand, computational implementations are found in the \texttt{changepoints} library \cite{KillickEckley2014} in \texttt{R} \cite{RcoreTeam2022} and the \texttt{SMOP} (Subset Multivariate Optimal Partitioning) library \cite{Pickering2016}, also in \texttt{R}. The latter is the one used to perform the comparisons with the method proposed in the paper. 

\begin{algorithm}
\caption{PELT Method \cite{KillickEtAl2012}}\label{alg:PELT}
 \hspace*{\algorithmicindent} \textbf{Input}: \\
	\hspace*{1cm} A set of data of the form, $(y_1, y_2, \ldots, y_T)$ where $y_i \in \mathrm{R}$.\\
 	\hspace*{1cm}	A set of measure $\mathcal{C}(\cdot)$ dependent on the data.\\
    \hspace*{1cm}	A penalty constant $\beta$ which does not depend on the number or location of change-points.\\
 	\hspace*{1cm}	A constant $K$ that satisfies (\ref{theorem:PELTTheorem}).
 	
 \hspace*{\algorithmicindent} \textbf{Initialize:} Let $T = $ length of data and set $F(0) = - \beta, \ cp(0) = NULL, R_1 = \{0\}$ \\
 \hspace*{\algorithmicindent} \textbf{Iterate:} for $\mathcal{T}^* = 1, 2, \ldots, T$
\begin{algorithmic}[1]
\State Calculate $F(\mathcal{T}^*) = \min_{\tau \in R_{\mathcal{T}^*}} [F(\mathcal{T}) + \mathcal{C}(y_{(\mathcal{T} + 1): \mathcal{T}^*}) + \beta]$
\State Let $\mathcal{T}^1 = arg \{\min_{\mathcal{T} \in R_{\mathcal{T}^*}} [F(\mathcal{T}) + \mathcal{C}(y_{(\mathcal{T}+1):\mathcal{T}^*}) + \beta]  \}$
\State Set $cp(\mathcal{T}^*) = cp[(\mathcal{T}^1), \mathcal{T}^1]$ 
\State Set $R_{\mathcal{T}^*  + 1} = \{ \mathcal{T} \in R_{\mathcal{T}^*} \cup \{ \mathcal{T}^* \}: F(\mathcal{T}) + \mathcal{C} (y_{(\mathcal{T} + 1: \mathcal{T}^*)}) + K \leq F(\mathcal{T}^*) \}$
\end{algorithmic}
\hspace*{\algorithmicindent} \textbf{Output}  the change points recorded in $cp(T)$
\end{algorithm}

\textbf{Cumulative Sum (CUSUM)\cite{Page1954}} \\

The CUSUM method is commonly used in quality control and, as its name indicates, makes use of cumulative sums to determine changes in a process parameter of interest; this parameter is called the quality value and is denoted by $\theta$. There are two versions of this method, a graphical one called $V$-mask and a tabular or algorithmic one, whose theoretical foundations can be found in \cite{Montgomery2008}.
	For the case of changes in the mean, let us again consider a time series $y_t, t = 1, 2, \ldots, T$. If the process is in control, each $y_t$ is assumed to have normal distribution of parameters $\mu_0, \ \sigma$. The parameter $\sigma$ is usually known or otherwise its estimator is used,
\begin{equation}
\hat{\sigma} = \sqrt{\frac{1}{T-1} \sum_{t = 1}^T(y_t - \mu_0)^2}
\label{eq:StdCUSUM}
\end{equation}

The CUSUM method in its tabular form works by accumulating values that deviate from $\mu_0$ above with the $C^{+}$ statistic and also accumulating deviations from $\mu_0$  below with the $C^{-}$ statistic. The $C^+$ and $C^-$  statistics are called the upper and lower CUSUM respectively and are calculated with the following expressions,

\begin{equation}
C_t^{+} = max \{0, y_t - (\mu_0 + K) + C_{t-1}^{+} \}
\end{equation}

\begin{equation}
C_t^{-} = max \{0, (\mu_0 - K) - y_t + C_{t-1}^{-} \}
\end{equation}

With $C_0^{+} = C_0^{-} = 0$. $K$ is usually called the reference or slack value and is usually chosen between $\mu_0$ and  $\mu_1$, the out-of-control value we are interested in detecting.
	Thus, if the change is expressed in standard deviation units as $\mu_1 = \mu_0 + \delta \sigma$  or ($\delta = |\mu_1 - \mu_0|/\delta$), then, $K$ is half the magnitude of the change or,

\begin{equation}
K = \frac{\delta}{2} \sigma = \frac{|\mu_1 - \mu_0|}{2}
\end{equation}
%

If $C_t^+$ or $C_t^-$ exceeds the decision interval, $H$, $y_t$ constitutes a change in the mean. A reasonable value for $H$ is five times the standard deviation of the series. \\

\textbf{Genetic Algorithm And Minimum Description Length \cite{Li2012}} \\

Again, a time series $\{y_t\}_{t = 1}^T$ is considered. On the other hand, we have as a function to optimize the Minimum Description Length (MDL) in its frequentist version,

\begin{equation}
MDL = - log(\mathcal{L}_{opt}) + \mathrm{Pen}
\end{equation}

Where $\mathcal{L}_{opt}$ is the maximum likelihood function of the observations, $y_t, t = 1, 2, \ldots, T$, after having replaced the parameters $\mathbf{\theta}$ by the estimators of the same type, $\hat{\mathbf{\theta}}_{ML}$ and $Pen$ refers to the penalty. For our case, it is of interest the MDL when $y_t \sim log-normal(\mu, \sigma)$, in the corresponding paper, the authors present the case of a series of observations with $Poisson(\lambda)$ distribution. Thus, if $y_t \sim log-normal(\mu, \sigma)$, its probability density function is given by,

$$f(y_t) = \frac{exp\{[ln(y_t) - \mu]^2 /2 \sigma^2 \}}{y_t \sigma \sqrt{2 \pi}}, \ y_t \geq 0, \ \mu \in \mathbb{R}, \ \sigma > 0$$


Assuming independence in the observations, the likelihood function of the whole time series is given by,

\begin{equation}
\mathcal{L}(\mu, \sigma^2) = \prod_{t = 1}^T f(y_t) = \frac{exp\{ -\frac{1}{2 \sigma^2} \sum_{t = 1}^T [ln(y_t) - \mu]^2 \}}{(\sigma \sqrt{2 \pi})^T (\prod_{t = 1}^T y_t) }
\end{equation}


The maximum likelihood estimates for the parameters are as follows,

$$\hat{\mu} = \frac{1}{T} \sum_{t = 1}^T ln (y_t)$$

$$\hat{\sigma}^2 = \frac{1}{T} \sum_{t = 1}^T [ln (y_t) - \hat{\mu}]^2$$


Then,

$$-ln(\mathcal{L}_{opt}) = \frac{T}{2} [1 + ln(2\pi) + ln(\hat{\sigma}^2)] + \sum_{t = 1}^T ln(y_t)$$


To include the influence of the change points, $\mu$ is estimated on a regime-by-regime basis. On the other hand, considering the change points $\mathcal{T} = \{\tau_1, \tau_2, \ldots, \tau_J\}$, the penalty is given by,

\begin{align}
\mathrm{Pen} &= \frac{3ln(T)}{2} + \sum_{j = 1}^{J+1} \frac{ln(\tau_j - \tau_{j-1})}{2} \nonumber \\
&+ ln(J) + \sum_{j = 2}^J ln(\tau_j)
\end{align}


Thus, the MDL takes the following form after omitting the values that do not depend on the change points,

\begin{align}
\label{eq:MDL}
MDL &= \frac{T}{2} ln(\hat{\sigma}^2) + \sum_{j = 1}^{J+1} \frac{ln(\tau_j - \tau_{j-1})}{2} \nonumber \\
&+ ln(J) + \sum_{j = 2}^J ln(\tau_j)
\end{align}


To optimize (\ref{eq:MDL}) a genetic algorithm is used, encoding the solutions as follows,

$$(J, \tau_0, \tau_1, \tau_2, \ldots, \tau_J, \tau_{J+1})$$


Where $\tau_0$ is time 1 and $\tau_{J+1}$ is time $T$. Thus each chromosome has $J+3$  elements. Then, with probability $p = 0.06$, $k$ initial solutions are created. This ratio is used as suggested by  \cite{Li2012}. Each solution is stored in a matrix named $Population_0$ with dimensions $k \times T$, such that the remaining $T  - (J+3)$ elements in each row are filled with zeros. Then, we proceed according to the pseudocode (\ref{alg:GA_MDL}).

\begin{algorithm}[H]
\caption{Multiple Change-point Detection Genetic Algorithm \cite{Li2012}}\label{alg:GA_MDL}
 \hspace*{\algorithmicindent} \textbf{Input}: \\
	\hspace*{0.5cm} $\cdot$ A set of $k$ initial solutions stored in $Population_0$ each one of the form, $Sol_i = (J, \tau_0, \tau_1, \tau_2, \ldots, \tau_J, \tau_{J+1}), \ i = 1, 2, \ldots, k$ with $\tau_j \in \{1, 2, \ldots, T\}$.\\
 	\hspace*{0.5cm}	$\cdot$ An MDL type objective function as (\ref{eq:MDL}) \\
 	
 \hspace*{\algorithmicindent} $\textbf{For} \ l = 1, 2, \ldots, r$\\
\begin{algorithmic}[1]
\State \hspace*{1cm} Calculate $MDL(S_i) \ \forall \ i = 1, 2, \ldots, k$
\State \hspace*{1cm} Sort every $MDL(Sol_i), \ i = 1, 2, \ldots, k$ with the worst solution in the first place and the best one in the last. Assign as rank 1, denoted $S_1$ to the worst solution and rank $k$, denoted $S_k$ to the worst one and select the $i-th$ solution with probability,

$$S_i/\sum_{j=1}^k S_j$$

as the Mother. The father is selected in the same manner with the $k-1$ remaining solutions.
\State \hspace*{1cm} Father, $Sol^{'} = (m, \delta_0, \delta_1, \delta_2, \ldots, \delta_m, \delta_{m+1})$ and mother, $Sol^{''} = (n, \eta_0, \eta_1, \eta_2, \ldots, \eta_n, \eta_{n+1})$, $\delta_0 = \eta_0 = Time \ 1$, $\delta_{m+1} = \eta_{n+1} = Time \ T$ are crossed in an uniform-like manner creating a new solution, $Child^{''} = (m+n, \tau_0, \tau_1, \tau_2, \ldots, \tau_{m+n}, \tau_{m+n+1})$. Duplicates are removed giving place to $Child^{''} = (J, \tau_0, \tau_1, \tau_2, \ldots, \tau_J, \tau_{J+1}), \ J \leq m+n$.
\State \hspace*{1cm} $Child^{''}$ is mutated adding up to each $\tau_j$ the result of throwing a three-sided dice with faces (-1, 0, 1) each one with probabilities, (0.3, 0.4, 0.3). Every element is stored in a chromosome, $Child'$.
\State \hspace*{1cm} Remove duplicate elements from $Child^{'}$ and store the times in a chromosome, $Child$ which in turn is stored in a new matrix $Population_l, \ l \geq 1$ until complete $k$ new solutions that replace the ones in $Population_{l-1}$. Every row in $Population_l$ should be different. 
\end{algorithmic}
\hspace*{\algorithmicindent} \textbf{END} $\textbf{For}$\\
\hspace*{\algorithmicindent} \textbf{Output:}  The best solution in the $r$ generations, $Sol_1^{*}, Sol_{2}^{*}, \ldots, Sol_r^{*}$ and the associated MDL values, $MDL(Sol_1^{*}), MDL(Sol_{2}^{*}), \ldots, MDL(Sol_r^{*})$
\end{algorithm}

\subsubsection{Experiments}

\textbf{PELT} \\

Since we know beforehand that $y_t \sim log-normal(\mu, \sigma)$, applying the concerning transformation, $x_t = log(y_t) \sim Normal(\mu, \sigma)$. Thus, one can assume a cost function $\mathcal{C}(\cdot) = \mathcal{L}(x_t)$, i.e., the likelihood of it, defined piece-wise

\begin{equation}
\mathcal{C}(y_{\tau_{j-1}: \tau_j}) = \sum_{j = 1}^J \mathcal{L}(y_{\tau_{j-1}: \tau_j}) = (2\pi \sigma^2)^{-T/2} exp \Bigg \{ \frac{-1}{2\sigma^2} \sum_{t = 1}^T (x_t - \mu_{r(t)})^2 \Bigg \}
\label{eq:PELTLikelihood}
\end{equation}

Where $\mu_{r(t)}$ is the regime bounded by the change-points $\tau_{j-1}, \tau_j$. On the other hand, we use the default definition for the penalty, the Akaike information criterion, AIC, whose expression is,

$$\texttt{Pen} = 2 \log(T)$$

Thus, we proceed to implement this method on the three data sets previously used for the proposed approach, that is, 1 change-point located at instant 825, two change-points at times 365 and 730 and three change-points at times 548, 823 and 973. The results are those shown in table (\ref{table:PELTResults1}).

\begin{table}[h]
	\begin{center}
		\begin{tabular}{c c c}
			\hline
			\textbf{Setting} & \textbf{Locations} &  \textbf{Detected Change-points by the algorithm}
			\\ \hline
			1 Change-point & $\mathcal{T} = \{825\}$ & $\hat{\mathcal{T}} = \{824\}$ \\
			\hline
			2 Change-points & $\mathcal{T} = \{365, 730 \}$ & $\hat{\mathcal{T}} = \{364, 729\}$ \\
			\hline
			3 Change-points & $\mathcal{T} = \{548, 823, 973\}$ & $\hat{\mathcal{T}} = \{546, 830, 972\}$ \\
			\hline                
		\end{tabular}
	\end{center}
	\caption{Results for first experiments with the PELT algorithm}
	\label{table:PELTResults1}
\end{table}

Although the results were accurate and the range of dispersion of each change-point detected by the algorithm to the real value is minimal, it should be noted that it was necessary to specify the distribution of the observations.

Implementing the same procedure by replacing $x_t$ by the original values, $y_t$ in (\ref{eq:PELTLikelihood}), we obtain the results given in (\ref{table:PELTResults2}).

\begin{table}[h]
	\begin{center}
		\begin{tabular}{c c p{2.5cm} p{2.5cm}}
			\hline
			\textbf{Setting} & \textbf{Locations} &  \textbf{Number of detected Change-points} & \textbf{Detected Change-points}
			\\ \hline
			1 Change-point & $\mathcal{T} = \{825\}$ & 86 & None \\
			\hline
			2 Change-points & $\mathcal{T} = \{365, 730 \}$ & 92 & None \\
			\hline
			3 Change-points & $\mathcal{T} = \{548, 823, 973\}$ & 93 & $\tau_3 = 973$ \\
			\hline                
		\end{tabular}
	\end{center}
	\caption{Results for second experiments with the PELT algorithm}
	\label{table:PELTResults2}
\end{table}

It can be seen that a deviation from distributional assumptions degrades the quality of the solutions; further comparisons could be carried out in future work to adapt the search method to include the proposed objective function, the Bayesian-MDL or using some non-parametric function. \\

\textbf{CUSUM} \\

To implement the CUSUM method, the \texttt{qcc} library \cite{Scrucca2004} of \texttt{R} \cite{RcoreTeam2022}  is used, and the function \texttt{cusum}($\cdot$). As the value of $K$ or the average distance magnitude between $\mu_1$ and $\mu_0$, $K = 1, 2, 3, 4$ are tested. In addition, $\mu_0$  is taken as the arithmetic mean of the observations, $\overline{Y} = \frac{1}{T}\sum_{t = 1}^T y_t$, where $\sigma$ is given by the expression (\ref{eq:StdCUSUM}) and as the decision interval, $H = 5$ is chosen. The results of the experiments are given in the table (\ref{table:CUSUMResults2}).

\begin{table}[h]
	\begin{center}
		\begin{tabular}{c c c p{2.5cm} p{2.5cm}}
			\hline
			\textbf{Setting} & \textbf{Locations} &  $K$ &  \textbf{Number of detected Change-points} & \textbf{Detected Change-points}
			\\ \hline
			&  & 1& 258 & None \\
			
			1 Change-point & $\mathcal{T} = \{825\}$ & 2& 230 & None \\
			&  & 3 & 0 & None \\ 
			&  & 4 & 0 & None \\ 
			\\ \hline
			&  & 1& 1051 & $\tau_1 = 365$ \\
			2 Change-points & $\mathcal{T} = \{365, 730\}$ & 2& 257 & None \\
			&  & 3 & 3 & None \\ 
			&  & 4 & 1 & None \\
			\\ \hline
			&  & 1& 888 & $\tau_1 = 548, \tau_3 = 973$ \\
			3 Change-points & $\mathcal{T} = \{548, 823, 973\}$ & 2& 119& None \\
			&  & 3 & 115 & None\\ 
			&  & 4 & 77 & None \\
			\\ \hline
		\end{tabular}
	\end{center}
	\caption{Results for experiments with the CUSUM method}
	\label{table:CUSUMResults2}
\end{table}

The method tends to detect by default, a large number of times as change-points. The best results are provided when taking $K = 1$ or when the distance between $\mu_0$ and $\mu_1$ is two units, however, this is due to the number of instants that are labeled as change-points, leading to an increase in the probability of including the true values of $\mathcal{T}$. \\

\textbf{Genetic Algorithm and MDL} \\

The genetic algorithm is implemented with the same characteristics as in the paper, 50 generations of 50 individuals each, a mutation rate of $3 \%$ and Gamma prior distributions for the parameters of the cumulative mean function, which in this case was also assumed to be Weibull. The results are given in the table (\ref{table:GAMDLResults}).

\begin{table}[h]
	\begin{center}
		\begin{tabular}{c c c}
			\hline
			\textbf{Setting} & \textbf{Locations} &  \textbf{Detected Change-points by the algorithm}
			\\ \hline
			1 Change-point & $\mathcal{T} = \{825\}$ & $\hat{\mathcal{T}} = \{100, 150\}$ \\
			\hline
			2 Change-points & $\mathcal{T} = \{365, 730 \}$ & $\hat{\mathcal{T}} = \{101, 152\}$ \\
			\hline
			3 Change-points & $\mathcal{T} = \{548, 823, 973\}$ & $\hat{\mathcal{T}} = \{93, 150\}$ \\
			\hline                
		\end{tabular}
	\end{center}
	\caption{Results for Genetic Algorithm with MDL}
	\label{table:GAMDLResults}
\end{table}

Although in all cases, the algorithm manages to detect a number of change-points that is not far from the real $J = 1, 2, 3$, in none of the three cases does it manage to capture the true values of $\mathcal{T}$. On the other hand, it was necessary to carry out an implementation according to the distribution of the data; in this case, previous information was available that allowed to approximate a $log-normal(\mu, \sigma)$ distribution, however, in complex cases where the parameter space may be larger or where the observations come from a mixture of distributions, it could operationally increase the definition of (\ref{eq:MDL}) and therefore, the detection of times of change. Therefore, the need to use a method that does not make assumptions regarding the behavior of the data, such as the one proposed in this work, becomes evident.

\subsection{Derivation of Objective Functions for other Intensities}\label{sec:DerivationObjFunctions}

Let's remember first the expressions for other intensity functions in addition to the Weibull one,

\begin{eqnarray}
	\label{eq2}
	\begin{array}{llll}
		\lambda^{(MO)}(t\mid\mathbf{\theta}) &=& \frac{\beta}{t+\alpha}, &\alpha ,\beta >0\\
		\lambda^{(GO)}(t\mid\mathbf{\theta}) &=& \alpha\beta\exp (-\beta t), &\alpha ,\beta >0\\
		\lambda^{(GGO)}(t\mid\mathbf{\theta}) &=& \alpha\beta\gamma t^{\gamma-1}\exp (-\beta t^\gamma), &\alpha ,\beta , \gamma >0.\\
	\end{array}
\end{eqnarray}

With each expression representing the Musa-Okumoto ($MO$) \cite{Musa84}, Goel-Okumoto ($GO$) and the Generalized Goel-Okumoto ($GGO$) \cite{Goel78} intensity functions. Next the respective mean cumulative function for each one are,

\begin{eqnarray}
	\label{eq1}
	\begin{array}{llll}
		m^{(MO)}(t\mid\mathbf{\theta}) &=& \beta\log \left(1+\frac{t}{\alpha}\right), &\alpha ,\beta >0\\
		m^{(GO)}(t\mid\mathbf{\theta}) &=& \alpha [1-\exp (-\beta t)], &\alpha ,\beta >0\\
		m^{(GGO)}(t\mid\mathbf{\theta}) &=& \alpha [1-\exp (-\beta t^\gamma)], &\alpha ,\beta , \gamma >0.\\
	\end{array}
\end{eqnarray}

The likelihood function under the presence of change-points and the independence of regimes takes the following form,

\begin{equation}
\label{eq11}
L(\mathbf{D}\mid\phi) \propto  \prod_{j=1}^{J+1} \left(e^{-[m(\tau_j\mid\theta_j)-m(\tau_{j-1} \mid \theta_j )]} \prod_{i=N_{\tau_{J-1}}+1}^{N_{\tau_j}} \lambda(d_i\mid\theta_j)   \right)  
\end{equation}

And after taking the logarithm of (\ref{eq11}) we have,

\begin{align}%
	\label{eq12}
	\log L(D \mid \phi) &= \left( \sum_{j=1}^{J+1}   m(\tau_{j-1}\mid\theta_j)-m(\tau_j\mid\theta_j)   \right) \nonumber\\
	 &+\left(\sum_{j=1}^{J+1} \quad \sum_{i= N_{\tau_{j-1}}+1}^{ N_{\tau_j}} \log\lambda(d_i\mid\theta_j) \right)\nonumber\\
	&=\sum_{j=1}^{J+1} \left(  m(\tau_{j-1}\mid\theta_j)-m(\tau_j\mid\theta_j)+\sum_{i= N_{\tau_{j-1}}+1}^{ N_{\tau_j}} \log\lambda(d_i\mid\theta_j)   \right)
\end{align}

The objective function, the Bayesian-MDL has the following structure,

\begin{align}
	\label{eq6}
MDL=-\log_2(L_{opt})+P.
\end{align}

With $P$ in (\ref{eq6}) being the penalty term,

\begin{equation}
\label{eq7}
P_{\tau}(\theta)=R\sum_{j=1}^{J+1} \frac{ln(\tau_1-\tau_{i-1})}{2} +ln(J) + \sum_{j=2}^{J}ln(\tau_j),    
\end{equation}

And for our case, (\ref{eq6}) becomes,

\begin{equation}\label{eq8}
\ln P_\tau(\theta) -  \ln L(D\mid\theta) - \ln f(\theta)    
\end{equation}

Finally let's remember the prior distribution for the two parameter model of the intensity function and the change-points,

\begin{align}
	\label{eq17}
	\begin{split}
\log f(\alpha,\beta, \tau_j) &\propto(\phi_{12}-1)\log\alpha - \phi_{11}\alpha \\
&+ (\phi_{22}-1)\log\beta - \phi_{21}\beta - log(T-1)
\end{split}
\end{align}

And also the case for the three-parameter model,

\begin{align}
	\label{eq18}
	\begin{split}
\log f(\alpha,\beta, \gamma, \tau_j) &= -\alpha \phi _{11}+\left(\phi _{12}-1\right) \log \alpha\\
&-\beta \phi _{21}+\left(\phi _{22}-1\right) \log \beta\\
&-\gamma \phi _{31}+\left(\phi _{32}-1\right) \log \gamma - log(T-1). 
\end{split}
\end{align}

Now we can proceed to ensemble the other objective functions.

\subsubsection{Musa-Okumoto (MO)}\label{sec:MODerivation}
Taking the expressions for the intensity function $\lambda^{(MO)}(t\mid\theta)$ and the cumulative mean function $m^{(MO)}(t\mid\theta)$ from (\ref{eq2}) and (\ref{eq1}) respectively, and replacing these values in (\ref{eq12}) we have,

\begin{equation}\label{eq21}
\begin{aligned}
	\log L(D\mid\phi)&=\sum_{j=1}^{J+1} \left(  m(\tau_{j-1}\mid\theta_j)-m(\tau_j\mid\theta_j)+\sum_{i= N_{\tau_{j-1}}+1}^{ N_{\tau_j}} \log\lambda(d_i\mid\theta_j)   \right)\\
	&=\sum_{j=1}^{J+1} \left(  \beta_j  \log \left(\frac{\alpha_j +\tau_{j-1}}{\alpha_j}\right) - \beta_j  \log \left(\frac{\alpha_j +\tau_j}{\alpha_j}\right) \right.  \\
	&\left. +\sum_{i= N_{\tau_{j-1}}+1}^{ N_{\tau_j}} \log\left( \frac{\beta_j }{\alpha_j +d_i}\right)   \right)\\
	&=\sum_{j=1}^{J+1} \left(  \beta_j  \log \left(\frac{\alpha_j +\tau_{j-1}}{\alpha_j}\right) - \beta_j  \log \left(\frac{\alpha_j +\tau_j}{\alpha_j}\right) \right. \\
	&\left. + (N_{\tau_j}-N_{\tau_{j-1}})\log(\beta_j) - \sum_{i= N_{\tau_{j-1}}+1}^{ N_{\tau_j}}\log(\alpha_j +d_i)   \right)\\
	&=\sum_{j=1}^{J+1}  \Bigg(\beta_j \left( \log \left(\alpha_j +\tau_{j-1}\right) -  \log \left(\alpha_j +\tau_j\right)\right) \\
	&+ (N_{\tau_j}-N_{\tau_{j-1}})\log(\beta_j) - \sum_{i= N_{\tau_{j-1}}+1}^{ N_{\tau_j}}\log(\alpha_j +d_i)\Bigg)
\end{aligned}
\end{equation}

Now, after replacing (\ref{eq7}), (\ref{eq21}), and (\ref{eq17}) in  (\ref{eq8}) we have, 

\begin{equation}
\label{eq22}
\begin{aligned}	
	P_{\tau}(\theta) - \ln f_\tau(D\mid\theta) - \ln f_\tau(\theta)
	&=2\sum_{i=1}^{J+1}\dfrac{\ln(\tau_i-\tau_{i-1})}{2}+  \ln(J) + \sum_{i=2}^J\ln(\tau_i)\\
	&-\sum_{j=1}^{J+1} \Bigg( \beta_j \left( \ln \left(\alpha_j +\tau_{j-1}\right) -  \ln \left(\alpha_j +\tau_j\right)\right)  \\
	&+ (N_{\tau_j}-N_{\tau_{j-1}})\ln(\beta_j)\\
	&-\sum_{i= N_{\tau_{j-1}}+1}^{ N_{\tau_j}}\ln(\alpha_j +d_i)  \Bigg)\\
	&-\sum_{j=1}^{J+1} \left((\phi_{12}-1)\ln\alpha_j - \phi_{11}\alpha_j \right.\\
	&\left.+ (\phi_{22}-1)\ln\beta_j - \phi_{21}\beta_j\right) + J \ \ln(T-1)\\
\end{aligned}
\end{equation}

\subsubsection{Goel-Okumoto (GO)}\label{sec:GODerivation}
As for  the previous case, we take the expressions  $\lambda^{(GO)}(t\mid\theta)$ and $m^{(GO)}(t\mid\theta)$ from (\ref{eq2}) and (\ref{eq1}) respectively, and replace these values in (\ref{eq8}). Then we have,

\begin{align}
		\label{eq23}
\log L(D\mid\phi)&=\sum_{j=1}^{J+1} \left(  m(\tau_{j-1}\mid\theta_j)-m(\tau_j\mid\theta_j)+\sum_{i= N_{\tau_{j-1}}+1}^{ N_{\tau_j}} \log\lambda(d_i\mid\theta_j)   \right)\nonumber\\
&=\sum_{j=1}^{J+1} \left(  \alpha_j\left[1- e^{-\beta_j \tau_{j-1}} \right] - \alpha_j\left[1- e^{-\beta_j \tau_{j}} \right] \right. \nonumber\\
&\left. +\sum_{i= N_{\tau_{j-1}}+1}^{ N_{\tau_j}} \log\left(\alpha\beta e^{-\beta d_i}\right)   \right)\nonumber\\
&=\sum_{j=1}^{J+1} \left(   \alpha_j\left[e^{-\beta_j \tau_{j}} - e^{-\beta_j \tau_{j-1}} \right] \right.\nonumber\\
&\left.+(N_{\tau_j}-N_{\tau_{j-1}})\log(\alpha\beta) -\beta\sum_{i= N_{\tau_{j-1}}+1}^{ N_{\tau_j}} d_i  \right)\nonumber\\
\end{align}

Replacing the expressions (\ref{eq7}), (\ref{eq23}), (\ref{eq17}) in the objective function of the expression (\ref{eq8}) we have that 

\begin{equation}\label{eq24}
\begin{aligned}
	P_{\tau}(\theta) - \ln f_\tau(D\mid\theta) - \ln f_\tau(\theta)
	&= 2\sum_{i=1}^{J+1}\dfrac{\ln(\tau_i-\tau_{i-1})}{2}+  \ln(J) + \sum_{i=2}^J\ln(\tau_i)\\
	&-\sum_{j=1}^{J+1} \Bigg(   \alpha_j\left[e^{-\beta_j \tau_{j}} - e^{-\beta_j \tau_{j-1}} \right] \\
	&+(N_{\tau_j}-N_{\tau_{j-1}})\ln(\alpha\beta) -\beta\sum_{i= N_{\tau_{j-1}}+1}^{ N_{\tau_j}} d_i  \Bigg)\\
	&-\sum_{j=1}^{J+1} \left((\phi_{12}-1)\ln\alpha_j - \phi_{11}\alpha_j \right. \\
	&\left. + (\phi_{22}-1)\ln\beta_j - \phi_{21}\beta_j\right) + J \ ln(T-1)
\end{aligned}
\end{equation}

\subsubsection{Generalized Goel-Okumoto (GGO)}\label{sec:GGODerivation}
Finally and once again, we take $\lambda^{(GGO)}(t\mid\theta)$ and $m^{(GGO)}(t\mid\theta)$ from (\ref{eq2}) and (\ref{eq1}) respectively, and replace these values in (\ref{eq12}) and then,

\begin{equation}\label{eq25}
\begin{aligned}
	\log L(D\mid\phi)&=\sum_{j=1}^{J+1} \left(  m(\tau_{j-1}\mid\theta_j)-m(\tau_j\mid\theta_j)+\sum_{i= N_{\tau_{j-1}}+1}^{ N_{\tau_j}} \log\lambda(d_i\mid\theta_j)   \right)\\
	&=\sum_{j=1}^{J+1} \Bigg(\alpha_j\left(1-  e^{-\beta_j \tau_{j-1}^{\gamma_j } }\right)- \alpha_j\left(1-  e^{-\beta_j \tau_{j}^{\gamma_j } }\right) \\
	& +\sum_{i= N_{\tau_{j-1}}+1}^{ N_{\tau_j}} \ln\left( \alpha_j\beta_j\gamma_j  d_i^{\gamma_j -1} e^{-\beta_j d_i^{\gamma_j}} \right)  \Bigg)\\
	&=\sum_{j=1}^{J+1} \Biggl(  \alpha_j\left(e^{-\beta_j \tau_{j}^{\gamma_j } }-  e^{-\beta_j \tau_{j-1}^{\gamma_j } }\right) + (N_{\tau_j}-N_{\tau_{j-1}}) \ln( \alpha_j\beta_j\gamma_j )\\
	&+ (\gamma_j -1)\Bigg( \sum_{i= N_{\tau_{j-1}}+1}^{ N_{\tau_j}}\ln d_i\Bigg)-\beta_j \Bigg( \sum_{i= N_{\tau_{j-1}}+1}^{ N_{\tau_j}}d_i^{\gamma_j}\Bigg)  \Biggr)
\end{aligned}
\end{equation}

Substituting (\ref{eq7}), (\ref{eq25}), and (\ref{eq18}) in (\ref{eq8}) we have, 

\begin{equation}\begin{aligned}
\label{eq26}
	P_{\tau}(\theta) - \ln f_\tau(D\mid\theta) - \ln f_\tau(\theta) 
	&=3\sum_{i=1}^{J+1}\dfrac{\ln(\tau_i-\tau_{i-1})}{2}+  \ln(J) + \sum_{i=2}^J\ln(\tau_i)\nonumber\\
	&-\sum_{j=1}^{J+1} \Bigg(  \alpha_j\left(e^{-\beta_j \tau_{j}^{\gamma_j } }-  e^{-\beta_j \tau_{j-1}^{\gamma_j } }\right) \nonumber \\
	&+ (N_{\tau_j}-N_{\tau_{j-1}}) \ln( \alpha_j\beta_j\gamma_j )\nonumber\\
	&+ (\gamma_j -1)\Bigg( \sum_{i= N_{\tau_{j-1}}+1}^{ N_{\tau_j}}\ln d_i\Bigg) \nonumber\\
	&-\beta_j \Bigg( \sum_{i= N_{\tau_{j-1}}+1}^{ N_{\tau_j}}d_i^{\gamma_j}\Bigg)  \Bigg)\nonumber\\
	&-\sum_{j=1}^{J+1} \left(\alpha_j \phi_{11}+\left(\phi _{12}-1\right) \ln (\alpha_j ) \right.\nonumber\\
	&\left.-\beta_j  \phi _{21}+\left(\phi_{22}-1\right) \ln (\beta_j )\right.\nonumber\\
	&\left.-\gamma_j  \phi _{31}+\left(\phi _{32}-1\right) \ln (\gamma_j )\right)+J\ln(T-1)
\end{aligned}
\end{equation}

\subsection{Assessing the influence of the number of Change-points on the algorithm performance}
\label{sec:NumberOfChangePointsExperiment}

\begin{figure}[h] 
\begin{center}
\includegraphics[width=1.1\textwidth]{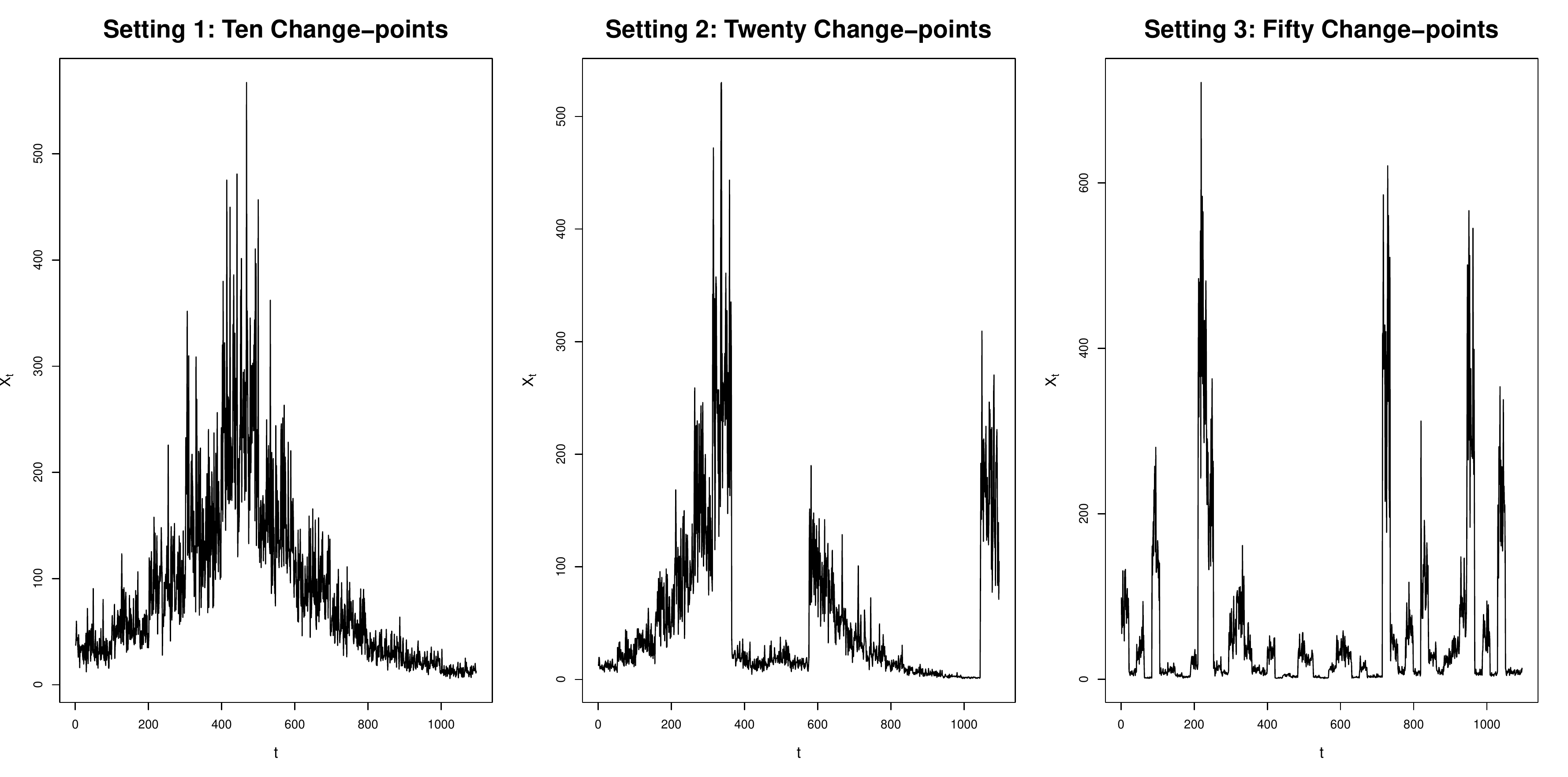}
\caption{Simulated time series behaviour - Second experiment}
\label{fig:SecondExperiment}
\end{center}
\end{figure}

In addition to the performance test of the algorithm proposed in this paper in section 4 (results and discussion), an additional test with a larger number of change-points is presented below. This time 10, 20 and 50 such instants are defined whose locations are given in the table (\ref{table:SimulationsTableJ_2}); as the number of change-points grows, locating them in the series becomes more complicated so it is chosen to divide the series into approximately equidistant lengths. The data are kept with a $log-normal$ distribution with the parameter $\mu$ taken randomly in the interval $[0.5, 6]$ and the parameter $\sigma = 0.32$ in all regimes (see Figure (\ref{fig:SecondExperiment})). Now, the purpose of these experiments is to determine whether there is a possible influence of the value of $J$ on the objective function and the results of the genetic algorithm as the penalty depends directly on this value and $T$, the length of the series (see, for example, expressions (\ref{eq22}), (\ref{eq24}) and (\ref{eq26})).

\begin{table}[h]
\begin{center}
\begin{tabular}{c p{7cm}}
\hline
\textbf{Number of Change-points} & \textbf{Locations}  \\ \hline
10                          & $\mathbf{\tau} = \{101, 201, 301, 401, 501, 597, 697, 797, 897, 997 \}$ \\
\hline
20                         & $\mathbf{\tau} = \{53, 105, 157, 209, 261, 313, 365, 417, 469, 525,$ $576, 629, 681, 731, 785, 837, 889, 941, 993, 1045\}$  \\
\hline
50                        & $\mathbf{\tau} = \{22, 43, 64, 85, 106, 127, 148, 169, 190, 211, 232, 253,$ $274, 295, 316, 337, 358, 379, 400, 421, 442, 463, 484, 505,$ $526, 547,568, 589, 610, 631, 652, 673, 694, 715, 736, 757,$ $778, 799, 820, 841, 862, 883, 904, 925, 946, 967, 988, 1009,$ $1030, 1051\}$  \\
\hline                
\end{tabular}
\end{center}
\caption{Different Change-points simulations considered - Second experiment}
\label{table:SimulationsTableJ_2}
\end{table}

\begin{figure}[h] 
\begin{center}
\includegraphics[width=1.1\textwidth]{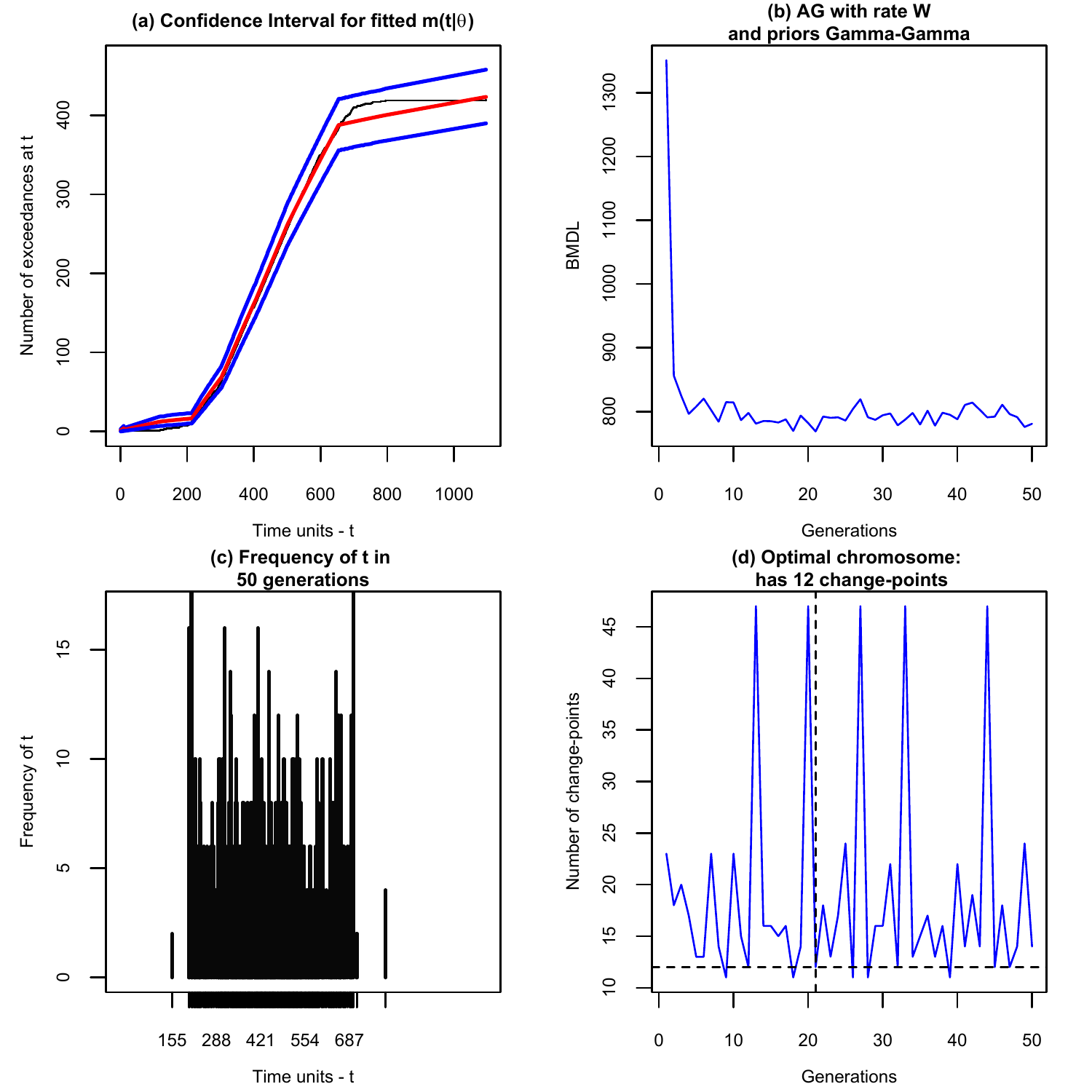}
\caption{First setting - second experiment}
\label{fig:10cpResults}
\end{center}
\end{figure}

\begin{figure}[h] 
\begin{center}
\includegraphics[width=1.1\textwidth]{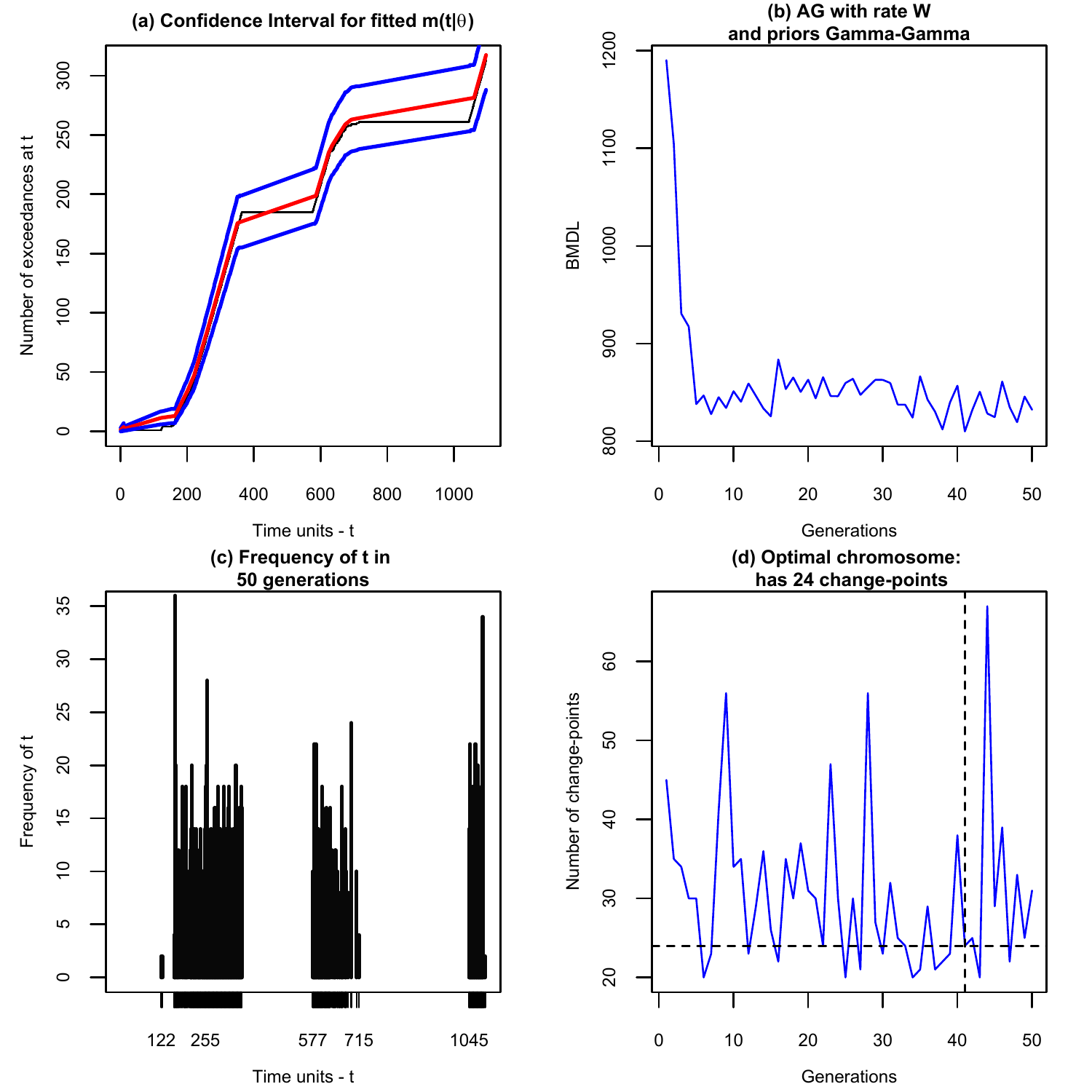}
\caption{Second setting - second experiment}
\label{fig:20cpResults}
\end{center}
\end{figure}

\begin{figure}[h] 
\begin{center}
\includegraphics[width=1.1\textwidth]{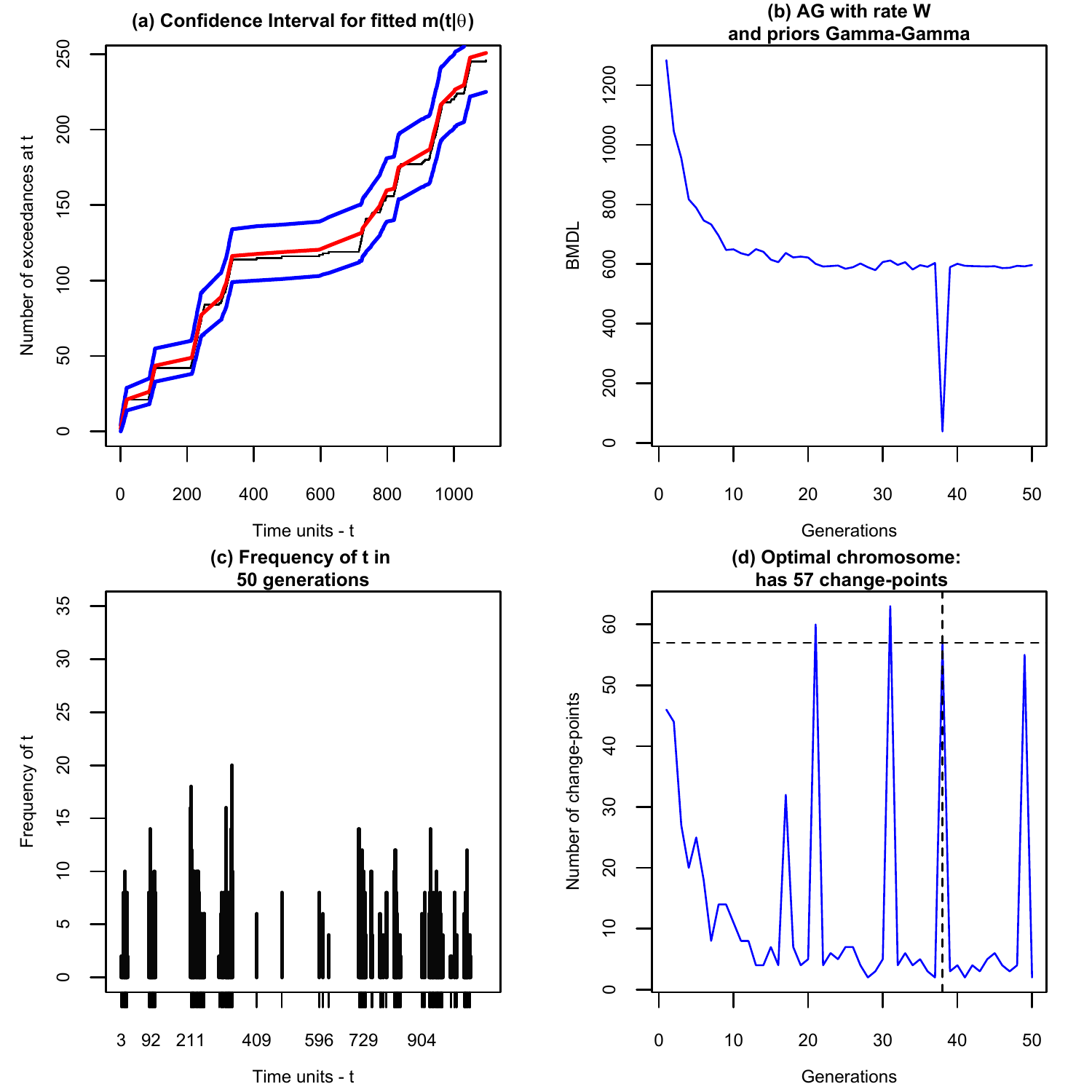}
\caption{Third setting - second experiment}
\label{fig:50cpResults}
\end{center}
\end{figure}

Implementing the genetic algorithm with 50 generations of 50 individuals each and taking the a priori distributions for a Weibull intensity function as in the previous cases, we have the results shown in Figures (\ref{fig:10cpResults}), (\ref{fig:20cpResults}) and (\ref{fig:50cpResults}) for $J = 10, 20$ and $50$, respectively.

Meanwhile, the table (\ref{table:SimulationsResultsExp2}) shows the results in relation to the minimum value of the objective function and where it is reached in each scenario. This is also shown in the graphs (d) of each of the figures according to the vertical dashed lines. On the other hand, graphs (a) in each case show a precise fit of the algorithm to the functions that determine the existence of the overshoots modeled from the cumulative mean function $m(t|\theta)$. Furthermore, it can be observed that in this implementation the presence of the change-points is captured, such that the estimate of $m(t|\theta)$ (red line) is very few units away from the actual values (black line). Likewise, the confidence intervals (blue lines) perfectly capture both the estimate of $m$ and its true value in all three cases.\\

Regarding the number of change-points, the genetic algorithm estimates values close to the true $J$, which can be seen in the three figures according to the graph (d) and the dashed horizontal line, as shown in the second column of the table (\ref{table:SimulationsResultsExp2}). For the first scenario, the algorithm estimates 2 additional change-points above the actual 10, in the second scenario it estimates 4 above the actual value and in the third scenario it also estimates 7 change-points above the true $J$.

Finally, in regards of the change-points detected correctly or approximately, for the first case these are the times  214, 303, 407 and 501 close to 201, 301, 401 and 501. In the second case the times 164, 201, 283, 351, 586, 624, 692 and 1060 are detected which are close to times  157, 209, 365, 576, 629, 681 and 1045 and finally in the third case the times 19, 87, 94, 104, 232, 242, 302,317, 335, 596, 723, 726, 778, 798, 820, 821, 834, 928, 949 and 960 which are close to the times 22, 85, 94, 106, 211, 232, 253, 295, 316, 337, 589, 715, 726, 778, 799, 820, 834, 925, 946 and 967.


Although there are moments not detected as change points when in fact they were, it should be noted that the estimation of the cumulative mean function is such that it allows forecasts to be made in relation to deviations from the mean and thus, of the moments where abrupt changes occur, which it also manages to capture. Furthermore, the purpose of the exercise is fulfilled by demonstrating that the algorithm is not sensitive to a large number of change points and is able to properly detect the number of change points and the true value of some of them.

\begin{table}[h]
\begin{center}
\begin{tabular}{c p{2.5cm} p{2.5cm} c}
\hline
\textbf{Setting} & \textbf{Detected number of Change-points} &  \textbf{Estimated Bayesian-MDL} & \textbf{Optimal Generation}
\\ \hline
10 Change-points & 12 & 768.70 & 21 \\
\hline
20 Change-points & 24 & 808.99 & 41 \\
\hline
50 Change-points & 57 & 38.43 & 38 \\
\hline                
\end{tabular}
\end{center}
\caption{Results for the Second experiment}
\label{table:SimulationsResultsExp2}
\end{table}

\end{document}